\documentclass[sn-nature]{sn-jnl}

\usepackage{graphicx}%
\usepackage{multirow}%
\usepackage{amsmath,amssymb,amsfonts}%
\usepackage{amsthm}%
\usepackage{mathrsfs}%
\usepackage[title]{appendix}%
\usepackage{xcolor}%
\usepackage{textcomp}%
\usepackage{manyfoot}%
\usepackage{booktabs}%
\usepackage{algorithm}%
\usepackage{algorithmicx}%
\usepackage{algpseudocode}%
\usepackage{listings}%

\usepackage{nicefrac}

\usepackage[T1, OT1]{fontenc}
\DeclareTextSymbolDefault{\dj}{T1}
\DeclareTextSymbolDefault{\DJ}{T1}

\begin{document}


\title{Scalable Feedback Stabilization of Quantum Light Sources on a CMOS Chip}



\author*[1]{Danielius~Kramnik}\email{kramnik@berkeley.edu}
\author*[2]{Imbert~Wang}\email{imbert@bu.edu}
\author*[3]{Anirudh~Ramesh}\email{anirudh@u.northwestern.edu}
\author[2]{Josep~M.~Fargas~Cabanillas}
\author[2]{\DJ{}or\dj{}e~Gluhovi\'{c}}
\author[1]{Sidney~Buchbinder}
\author[1]{Panagiotis~Zarkos}
\author[1]{Christos~Adamopoulos}
\author[3]{Prem~Kumar}
\author[1]{Vladimir~M.~Stojanovi\'{c}}
\author[2]{Milo\v{s}~A.~Popovi\'{c}}

\affil[1]{\orgdiv{Department of Electrical Engineering and Computer Sciences}, \orgname{University of California Berkeley}, \orgaddress{\city{Berkeley}, \state{CA}  \postcode{94709}, \country{USA}}}

\affil[2]{\orgdiv{Department of Electrical and Computer Engineering and Photonics Center}, \orgname{Boston University}, \orgaddress{\city{Boston}, \state{MA}  \postcode{02215}, \country{USA}}}

\affil[3]{\orgdiv{Center for Photonic Communication and Computing, Department of Electrical and Computer Engineering}, \orgname{Northwestern University}, \orgaddress{\city{Evanston}, \state{IL}  \postcode{60208}, \country{USA}}}

\abstract{
Silicon photonics is a leading platform for realizing the vast numbers of physical qubits needed to achieve useful quantum information processing because it leverages mature complementary metal-oxide-semiconductor (CMOS) manufacturing to integrate on-chip thousands of optical devices for generating and manipulating quantum states of light. 
A challenge to the practical operation and scale-up of silicon quantum-photonic integrated circuits, however, is the need to control their extreme sensitivity to process and temperature variations, free-carrier and self-heating nonlinearities, and thermal crosstalk.
To date these challenges have been partially addressed using bulky off-chip electronics, sacrificing many benefits of a chip-scale platform and limiting the practically achievable system size.
Here, we demonstrate the first electronic-photonic quantum system-on-chip (EPQSoC) consisting of quantum-correlated photon-pair sources stabilized via on-chip feedback control circuits, which is fabricated in a high-volume 45\,nm CMOS microelectronics foundry.
We use non-invasive photocurrent sensing in a tunable microring cavity photon-pair source to actively lock it to a fixed-wavelength pump laser while operating in the quantum regime, enabling large scale microring-based quantum systems.
This is the first demonstration of such a capability, achieving a high coincidences-to-accidentals ratio of $\sim$134 with an ultra-low $g^{(2)}(0)$ of 0.021 at $\sim$2.2\,kHz off-chip detected pair rate and $\sim$3.3\,MHz/mW$^2$ on-chip pair generation efficiency, and over 100\,kHz off-chip detected pair rate at higher pump powers ($\sim$1.5\,MHz on-chip). 
We show that these sources maintain stable quantum properties in the presence of temperature variations and operate reliably in a practical setting with many adjacent photon-pair sources creating thermal disturbances on the same chip.
Such dense integration of electronics and photonics enables implementation and control of quantum-photonic systems at the scale needed to achieve useful quantum information processing with CMOS-fabricated chips.
}

\maketitle
\newpage


\section*{Introduction}\label{introduction}

Photonics is a compelling approach for the development of quantum technologies such as quantum computing, secure communication, and sensing due to its robustness to decoherence at room temperature, natural compatibility with optical interconnects for entanglement distribution, and ability to be miniaturized into chip-scale devices \cite{rudolph2017optimistic, wang2020integrated}.
Silicon photonics \cite{adcock2020advances} offers the most scalable platform for quantum-photonic systems, enabling them to be built using mature semiconductor fabrication techniques developed in the complementary metal-oxide-semiconductor (CMOS) microelectronics industry \cite{giewont2019globalwgprocess, rakowski2020cloprocess}, which routinely produces chips with billions of functioning transistors in high volumes.
Within this platform, third-order ($\chi^{(3)}$) optical nonlinearity in silicon and silicon nitride waveguides \cite{sharping2006siliconfwm, harada2008entangled} and microring resonators \cite{clemmen2009microringfwm} enables the generation of quantum-correlated photon pairs via spontaneous four-wave mixing (SFWM), which can be used to build compact sources of heralded flying qubits.
Furthermore, recent advances in heterogeneous material integration onto silicon-photonic chips have the potential to create a library of optical devices needed to assemble a large-scale photonic quantum computer in a CMOS foundry platform \cite{alexander2024manufacturable}. 
These include III-V lasers for pumping single-photon qubit sources \cite{bian2023clolaser, mahmudlu2023fwmsourcelaserintegration}, deterministic quantum emitters based on III-V quantum dots \cite{kim2017hybrid, katsumi2019quantum, larocque2024tunable}, electro-optic materials for high-speed, low-loss optical switches and modulators used in state preparation and manipulation \cite{eltes2019batio,eltes2020integrated,wang2022linbo3modulatorintegration, wang2023integrated}, and high-efficiency superconducting nanowire single-photon detectors (SNSPDs) for qubit readout and feedforward operations \cite{pernice2012snspdsiph, akhlaghi2015waveguide, najafi2015chip}.
To function properly, however, silicon quantum photonic devices, especially micro-resonator based ones like photon-pair sources and filters, require continuous monitoring and control by electronic circuits.
To date this has been carried out using bulky off-chip electronics \cite{carolan2019scalablecontrol}, limiting the practically achievable system size and complexity.
Realizing the full potential of silicon photonics as a platform for quantum information processing still requires this classical control bottleneck to be resolved.

Microring resonator photon-pair sources are an essential building block for silicon quantum photonics 
as the field enhancement from their high cavity Q-factors and finesse boosts the rate and spectral brightness with which photon-pairs are generated \cite{caspani2017integrated}, replacing millimeter-scale photon sources based on straight waveguides with resonators as small as a few tens of \textmu m in diameter \cite{savanier2016photon}.
Such miniaturized sources can provide the vast numbers of heralded single photon qubits needed per chip for useful quantum information processing.
However, the high (typically $\gg$10,000) Q-factors of microrings optimized for SFWM also create calibration and control challenges that have limited practical quantum-optical experiments based on silicon photonics to no more than a few interacting microring sources \cite{silverstone2015qubit, faruque2018chip, lu2020threedim, llewellyn2020teleportationchip, arrazola2021xanadu}, with recent larger systems so far still relying on non-resonant devices \cite{wang2018multidimensional, bao2023vlsiquantumphotonics}.
These challenges stem from the need for the narrow-linewidth resonances of all microrings in a system to be closely aligned to the same set of pump laser and output wavelengths (historically called the ``signal'' and ``idler'') to produce streams of heralded photons capable of quantum interference \cite{faruque2018chip, yard2022chip}.

The sources of wavelength mismatch between different microring resonators can broadly be separated into two types: static mismatches and dynamic effects.
The former arise from fabrication-induced variations inherent to semiconductor manufacturing such as uneven device layer film thicknesses and etch depths across a wafer, leading to initial differences on the order of nanometers between microring resonance wavelengths even in state of the art processes \cite{horikawa2017resonant, jayatilleka2021processvariation}. 
Dynamic effects, however, arise from factors that are not fixed at fabrication time such as the temperature and free carrier concentration in each microring.
These include thermal crosstalk from the calibration and reconfiguration of other nearby photonic components over the course of operating a system, as well as optical loss mechanisms such as two-photon absorption (TPA) \cite{dinu2003sitpabeta} and free carrier absorption (FCA) \cite{soref1987electrooptical}, which lead to self-heating as the circulating optical power increases. 
When the pump laser power is boosted to maximize the photon-pair generation rate, this self-heating creates a positive feedback loop that eventually manifests as thermal hysteresis \cite{almeida2004opticalbistability, priem2005opticalbistability}, preventing an on-resonance lock between a microring and pump laser unless the microring's temperature is initialized and stabilized properly \cite{sun2016bitstatistics, guo2019nonclassical}.
Additionally, plasma dispersion from electrons and holes generated by TPA and other parasitic absorption mechanisms shifts microring resonances in the opposite direction of self-heating \cite{xu2006carrier}.
This can result in unstable behaviour such as optical pulsing \cite{johnson2006opticalpulsing} where the two effects take turns pushing the microring's resonances in opposite directions and cause the cavity's optical energy to oscillate, sometimes chaotically \cite{chen2012bistability}, which prevents stable photon-pair generation.
To make microring-based sources viable for large-scale quantum systems-on-chip, these calibration and control issues must be addressed using chip-scale electronics that are densely co-packaged with the silicon photonics and capable of correcting resonance wavelength errors in real time without disrupting the quantum characteristics of the system.

\begin{figure}[h!]
    \centering
    \includegraphics[width=\textwidth]{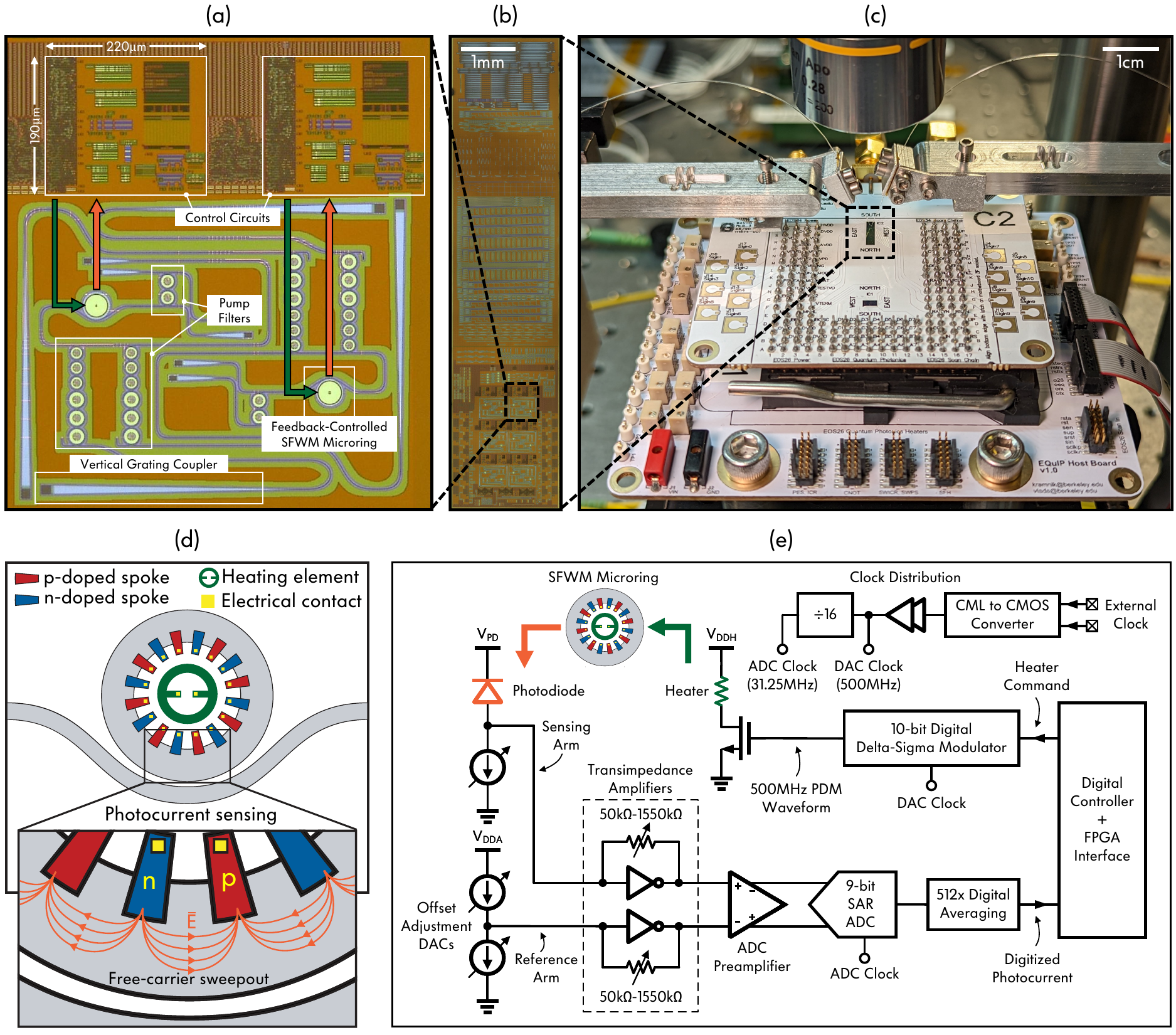}
    \caption{\textbf{Electronic-photonic-quantum system-on-chip:}  (a) Micrograph of two interleaved electronic-photonic quantum circuit blocks with 220\,\textmu m\,$\times$\,190\,\textmu m control circuits and 19.1\,\textmu m radius SFWM microrings. (b) Micrograph showing the entire $2\,\text{mm} \times 9\,\text{mm}$ electronic-photonic CMOS chip, which contains 12 photon-pair source blocks towards the bottom. (c) The CMOS chip is flip-chip bonded to a printed circuit board that provides power and interfaces to an FPGA and PC. The silicon handle is removed via a XeF$_2$ dry etch to enable optical access from the back side of the chip via lensed fiber probes coupling to the VGCs from above. (d) Diagram of the SFWM-optimized microring with \textit{p-i-n} diodes for carrier sweepout in reverse bias. (e) Schematic of the feedback-controlled SFWM microring pair source, showing the on-chip circuits for thermal tuning and in-cavity light intensity monitoring via the \textit{p-i-n} junction photodiodes.}
    \label{fig1}
\end{figure}

To address this need, we introduce the first electronic-photonic quantum system-on-chip that is capable of stabilizing SFWM in microring resonators in a truly \textit{in-situ} manner by sensing and minimizing their wavelength errors in the quantum regime using on-chip electronics with a compact footprint of 220\,\textmu m\,$\times$\,190\,\textmu m.
Figs. \ref{fig1}(a-c) show micrographs of the system-on-chip, where active silicon photonics optimized for generating photon-pairs in the C-band (1530-1565\,nm wavelength) are fabricated in GlobalFoundries' 45RFSOI CMOS process using the same silicon-on-insulator (SOI) layer as nearby analog and digital control electronics \cite{stojanovic2018monolithic}.
This approach was originally developed for integrating classical optical interconnects with microprocessors, where the operation of hundreds of silicon-photonic devices alongside tens of millions of transistors has already been demonstrated \cite{orcutt2012open, sun2015nature, sun2020teraphy, wade2020teraphy, wade2021teraphy}.
We also previously demonstrated passive quantum photonics in this platform \cite{gentry2015quantum}, including a single chip with SFWM photon-pair generation and $>$95\,\text{dB} pump filtering \cite{gentry2018monolithic}, as proofs of concept that quantum photonics can be implemented in a CMOS platform supporting electronic circuits.
In this paper, we introduce new techniques to enable scalable quantum photonics to be integrated within this platform.
We integrate non-invasive photocurrent sensors within SFWM-optimized microrings that allow their circulating optical power to be sensed directly, avoiding the need to switch between classical detectors for calibration and quantum detectors for operation \cite{carolan2019scalablecontrol}, or to use separate post-filters and photodetectors that occupy additional area and must themselves be aligned to the correct range of wavelengths \cite{llewellyn2020teleportationchip}.
This demonstration of a fully integrated, modular control approach paves the way for silicon quantum photonics to achieve the massive scale required for future generations of quantum information systems.

\section*{Results}

\subsubsection*{{\small Electronic-Photonic Quantum System-on-Chip}}

Figs. \ref{fig1}(d-e) present schematic diagrams of the system for stabilizing SFWM in silicon microring resonators.
Embedded sensors allow integrated control electronics to monitor the optical power in a resonator and then thermally tune it via an integrated heater to lock to a pump laser.
Twelve such systems with variations in the photonic device parameters are present in the fabricated chip, utilizing 400\,nm wide single-mode waveguides patterned in the sub-100\,nm thick SOI device layer to confine and route light.
Optical coupling to the chip is achieved via bi-layer unidirectional vertical grating couplers (VGCs) implemented in the silicon body and polysilicon gate layers \cite{wade2015grating, notaros2016grating}.
At the core of each system is a high-Q microring resonator of radius $\sim$20\textmu m designed to produce photon-pairs via SFWM, enable optical power monitoring within the cavity, and support resonance tuning.
Since only one thickness of SOI is available in standard CMOS microelectronics processes and the doped polysilicon gate layer introduces large losses in microrings \cite{atabaki2016polysilicon}, a conventional rib waveguide geometry was avoided.
Instead, optical power sensing is achieved using a wide multimode SOI waveguide with interdigitated \textit{p-i-n} diodes along the inner edge of the microring away from the fundamental optical mode.
This establishes multiple parallel photodiodes that sweep out and collect photogenerated carriers in the silicon waveguide.
An adiabatically wrapped ring-bus coupler prevents higher-order transverse modes from being excited in the 2.8\,\textmu m wide waveguide cross-section of the resonator, coupling only to the fundamental mode propagating closer to the outer edge of the ring \cite{hosseini2010systematic, shainline2013depletion}.
This separation between heavily doped contacts and propagating light allows the resonator to maintain an intrinsic Q on the order of $10^5$, comparable to undoped single-mode microrings in the same process \cite{orcutt2012open, gentry2015quantum}.
The ring radius and waveguide width have been selected to optimize the four-wave mixing strength within the constraints of the CMOS platform by trading off between the mode volume, intrinsic Q-factor, and dispersion (see Supplementary Information \S 1).

Although silicon is transparent in the C-band, carriers can still be generated by defect and surface state absorption \cite{grillanda2015ssa}, as well as TPA at higher circulating optical powers in the resonator \cite{dinu2003sitpabeta}.
These effects produce a responsivity on the order of 1\,mA/W of bus-waveguide pump light that is sufficient for sensing the alignment of the resonance wavelength to the pump laser by on-chip circuits, while avoiding the loss penalty and intrinsic Q reduction of a drop port photodetector or absorbing material such as SiGe \cite{alloatti2016resonance}.
Carrier sweepout also mitigates carrier-induced losses and nonlinear effects.
This boosts the efficiency of the four-wave mixing process \cite{ong2013efficient, savanier2015optimizing} and delays or eliminates the onset of free-carrier-induced instability, with the tradeoff being that the work done to extract the carriers increases self-heating and lowers the pump power threshold for optical bistability \cite{gray2020thermo}. 
To match the photocurrent produced at typical pump powers of \mbox{100\,\textmu W~--~1\,mW} to CMOS voltage levels, we use an inverter-based transimpedance amplifier (TIA) with adjustable gain of \mbox{50\,k$\Omega$~--~750\,k$\Omega$} to amplify the signal before it is digitized by an on-chip analog-to-digital converter (ADC) and sent to a digital circuit block for processing. 
A pseudo-differential amplifier structure is implemented to improve power supply noise rejection and programmable current sources at the input enable automatic offset calibration.
Further circuit implementation and calibration details are provided in Supplementary Information \S 2. 

\begin{figure}[h!]
    \centering
    \includegraphics[width=\textwidth]{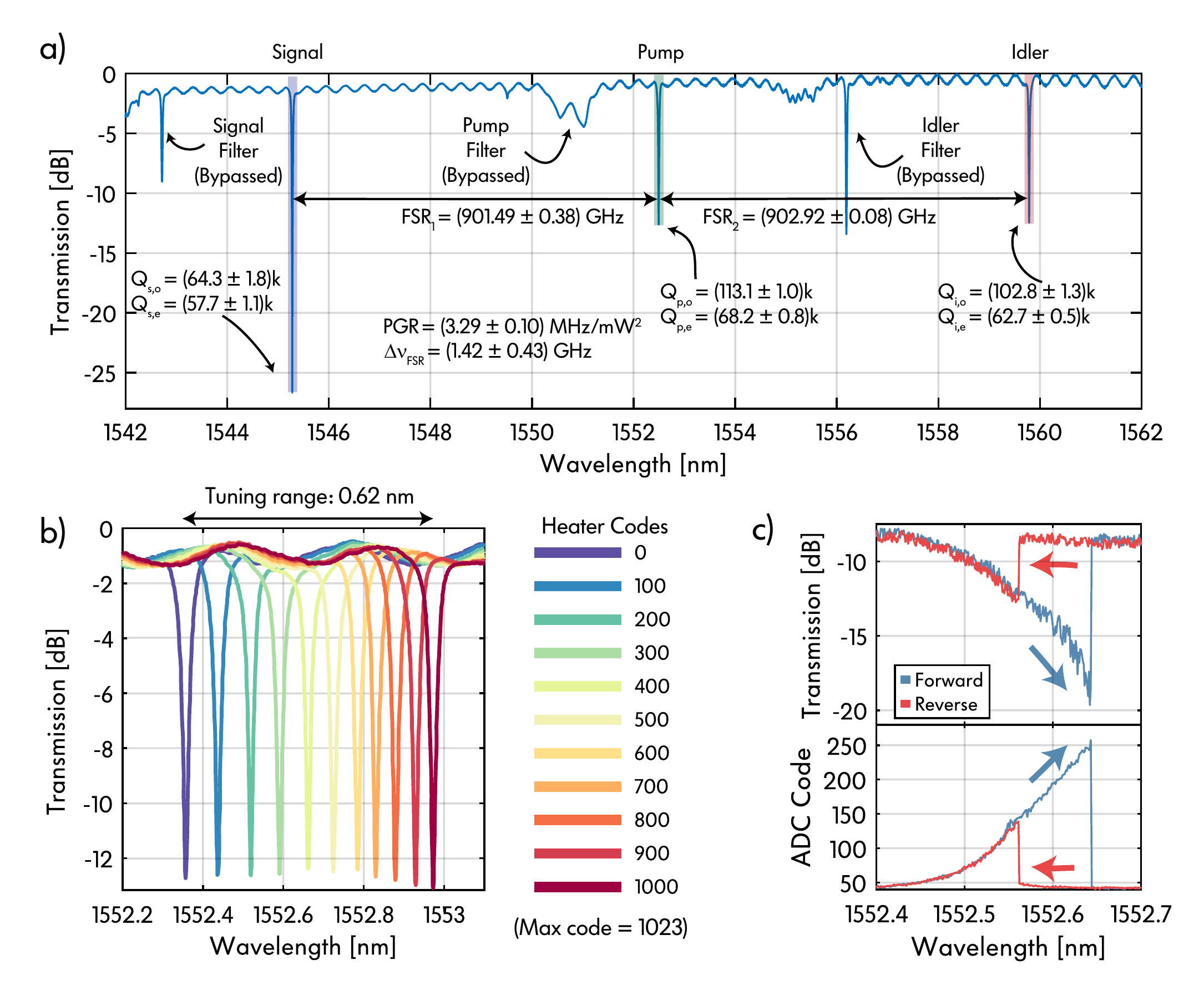}
    \caption{\textbf{Classical characterization of the electronic-photonic integrated circuit block:}  (a) Normalized transmission spectrum of the SFWM microring system site with fitted intrinsic and extrinsic Q-factors, FSR mismatch, and predicted PGR based on the equations in Supplementary Information \S 1. The three resonances at 1545.3\,nm (signal), 1552.5\,nm (pump), and 1559.8\,nm (idler) are chosen for photon-pair generation. On-chip pump ASE cleanup and pump rejection filters are bypassed for the experiments reported here.
    (b) Standalone characterization of the SFWM microring heater DAC with the $V_\text{DDH}$ supply set to $1.3\,\text{V}$, which produces a least significant bit (LSB) step of 0.6\,pm (76\,MHz) and overall wavelength tuning range of 0.62\,nm with 5.1\,mW maximum heater power.
    (c) Normalized transmission spectrum (top plot) and the corresponding digitized photocurrent from the SFWM microring (bottom plot) with $-5.1\,\text{dBm}$ on-chip pump power as the laser is swept forward from short to long wavelengths and in reverse from long to short wavelengths, demonstrating the thermal bistability in the microring. The microring resonance must be initialized to a longer wavelength than the pump laser (corresponding to the forward sweep) to make it possible to access the fully on-resonance state.} 
    \label{fig2}
\end{figure}

The integrated digital circuits can either run a fully self-contained hard-coded algorithm to search for and lock near the maximum photocurrent or send the ADC data to an external controller and receive commands through a serial scan chain.
Once this digital system processes the ADC measurements to generate a feedback command, resonance wavelength adjustments are executed by an on-chip digital-to-analog converter (DAC) driving a resistive heater in the center of the ring, which tunes the resonance wavelengths via the thermo-optic effect in silicon \cite{xie2020thermally}.
We used a high-frequency (500\,MHz) switching output stage driven by a delta-sigma modulator to improve the circuit efficiency and ensure that the 10-bit DAC command to heater power transfer curve is monotonic \cite{sun2013pdm}, which is required to create a stable negative feedback loop to control the resonance wavelength.
Adjusting an externally-provided DAC output stage supply voltage ($V_\text{DDH}$) allows the tuning range and least significant bit (LSB) step size to be scaled together proportionately -- for the experiments here we selected $V_\text{DDH} = 1.3\,\text{V}$, resulting in an LSB step of 0.6\,pm (76\,MHz) and overall wavelength tuning range of 0.62\,nm with 5.1\,mW maximum heater power.
The LSB step is chosen to be small enough compared to the $\sim$5\,GHz full-width half-maximum (FWHM) linewidths of the cold SFWM microring resonance to enable locking to within a few percent of the peak photocurrent, maximizing the pair generation rate.
Fig. \ref{fig2} shows the transmission spectra of a microring photon-pair source with varying DAC settings applied to the heater, and photocurrent measurements by the ADC as the heater is swept from cold to hot and vice versa, displaying the characteristic thermal bistability of high-Q silicon microrings.

\subsubsection*{{\small Feedback Stabilization of Photon-Pair Generation in a Microring}}

\begin{figure}[h!]
    \centering
    \includegraphics[width=\textwidth]{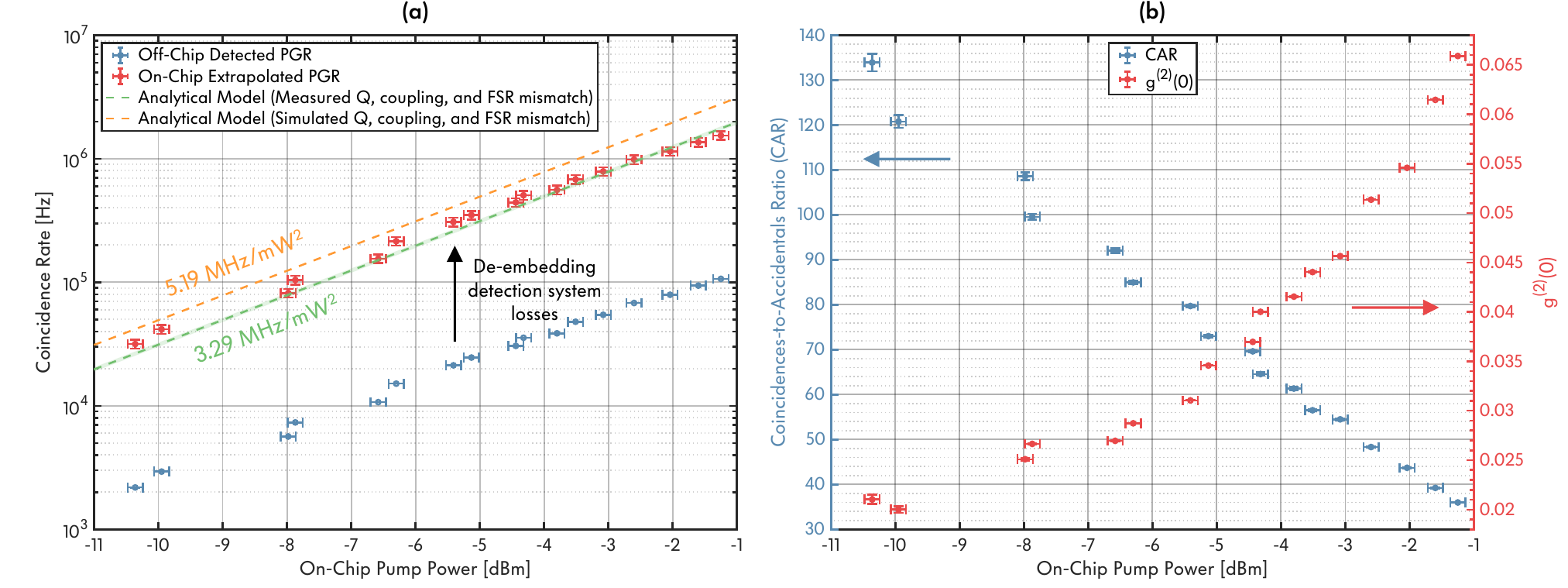}
    \caption{\textbf{Quantum characterization of the SFWM photon-pair source with real-time feedback control:}
    (a)  Plot of the coincidence rate detected at the SNSPDs and extrapolated on-chip pair generation rate (PGR) under feedback control versus the on-chip pump power at the input of the SFWM microring.
    Expected trends calculated using the analytical model described in Supplementary Information \S 1 are also plotted.
    The target design has $Q_{o,(\text{s,p,i})} = 116.5 \times 10^3$, $Q_{e,(\text{s,p,i})} = 87.4 \times 10^3$, and FSR mismatch of 1.98\,GHz, yielding a predicted PGR efficiency of $5.19\,\text{MHz}/\text{mW}^2$.
    The fabricated microring on the tested die has parameters given in Fig. \ref{fig2}(a), yielding a predicted PGR efficiency of $3.29 \pm 0.13\,\text{MHz}/\text{mW}^2$, closely matching the extrapolated on-chip PGR.
    (b)  Plot of the coincidences-to-accidentals ratio (CAR) and $g^{(2)}(0)$ versus on-chip pump power.
    The maximum measured CAR is $133.9 \pm 1.9$ and corresponding $g^{(2)}(0)$ is $0.021$ at $2.2\,\text{kHz}$ off-chip coincidence rate ($\approx 31.7\,\text{kHz}$ estimated on-chip PGR) with $-10.4\,\text{dBm}$ on-chip pump power assuming equal losses at each grating coupler ($3.0 \pm 0.1\,\text{dB}$), which results in a total loss between SFWM microring and coincidence detection event of $14.5 \pm 0.2 \,\text{dB}$ when accounting for all photon loss mechanisms in the three SNSPD channels.
    The signal photons are sent through a 50/50 beamsplitter to enable $g^{(2)}$ measurements, and the detected off-chip coincidence rate includes counts from both signal channels -- the full experimental setup is illustrated in Fig. \ref{fig3}.
    The maximum detected coincidence count rate is $106.6\,\text{kHz}$ ($\approx$$1.54\,\text{MHz}$ estimated on-chip PGR) with a CAR of $36.0 \pm 0.1$ and $g^{(2)}(0)$ of $0.066$ at an on-chip pump power of $-1.2\,\text{dBm}$.
    We use a 320\,ps wide coincidence window ($\pm 2 \sigma$) for all measurements reported here.
    All errors bars represent the standard deviations of measured quantities.
    }
    \label{fig3}
\end{figure}

In order to access the state where its resonance is aligned with the pump laser, the microring has to be initialized hot to place the resonance at a longer wavelength than the pump laser.
Then, as it is cooled, its resonance shifts to shorter wavelengths until at a certain point the additional self-heating from absorbing some of the pump light counteracts this by exactly the right amount to align the two, inducing the maximum amount of photocurrent in the \textit{p-i-n} diodes.
Any further reduction in heater power beyond this point causes the resonance to quickly snap away to shorter wavelengths as the self-heating effect reverses from stable negative feedback to unstable positive feedback \cite{sun2016bitstatistics}.
The responsivity of the microring, coupling condition, pump laser optical path insertion losses, and sensing circuit gain are all subject to fabrication-induced variations, making it unfeasible to predict the ADC reading where the microring balances on the edge of instability.
Therefore, we perform a calibration step before attempting to lock the microring to the pump laser, separating the feedback control scheme into three stages.
First, a hot-to-cold sweep determines the ADC readings corresponding to the dark current and maximum photocurrent.
Second, a quick reset to the hot state re-initializes the microring properly and a new hot-to-cold sweep brings it to the point where the photocurrent again starts increasing.
Third, a proportional-integral (PI) controller engages and regulates the photocurrent to just below the maximum value (95\,\% of the way from the dark current to maximum photocurrent levels in the experiments reported here) by fine-tuning the heater DAC code, balancing the resonance on the edge of instability while leaving enough margin to reject thermal disturbances and small variations in pump power from fiber alignment drift.
A small deadband around the regulation point eliminates spurious dithering of the DAC code, allowing it to settle to a constant value.
At this point, quantum-correlated photon-pairs are generated via SFWM in the microring, separated from residual pump light using off-chip bandpass filters, and detected using a commercial SNSPD system with time-correlated single-photon counting (TCSPC).
Fig. \ref{fig3}(a) shows the raw detected photon-pair coincidence rate and the extrapolated on-chip pair generation rate (PGR) after de-embedding insertion losses and SNSPD detection efficiencies, and Fig. \ref{fig3}(b) shows coincidences-to-accidentals ratio (CAR, which is the quantum signal to noise ratio) and the second order correlation function $g^{(2)}(0)$ at the SNSPDs, demonstrating the single-photon nature of heralded photons from the source.
We characterized the ability of the control system to stabilize the SFWM microring over approximately an order of magnitude in pump power ranging from $-10.4\,\text{dBm}$ to $-1.2\,\text{dBm}$ on-chip, allowing the PGR-CAR tradeoff to be tuned to the requirements of a given quantum system.

\begin{figure}[h!]
    \centering
    \includegraphics[width=\textwidth]{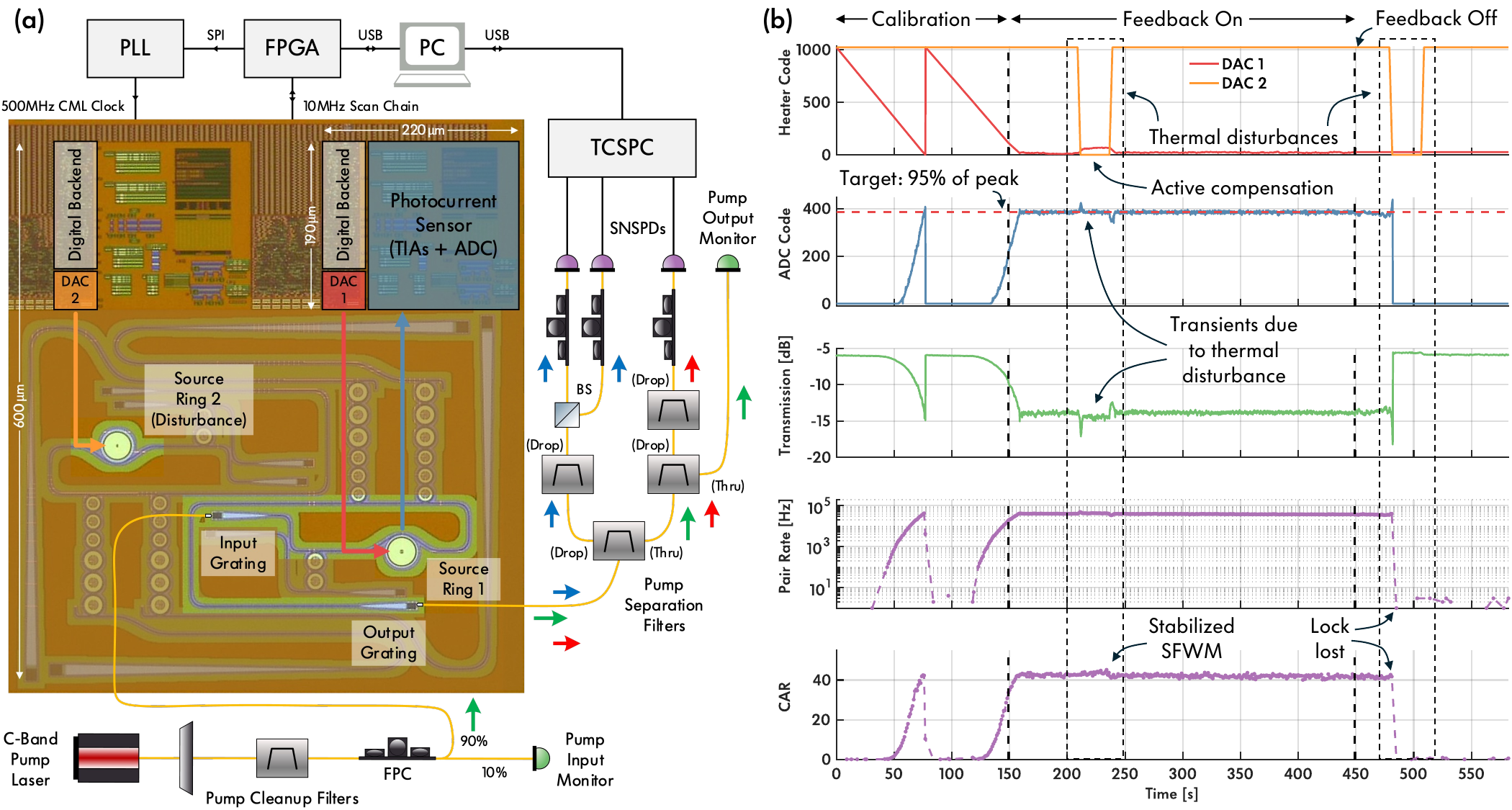}
    \caption{\textbf{Experimental demonstration of feedback-controlled SFWM under thermal disturbances:}  (a) Schematics of the experimental setup with active feedback control and off-chip pump filtering, highlighting active devices on the chip. A 50/50 beamsplitter added to the signal channel enables the $g^{(2)}(0)$ measurements shown in Fig. \ref{fig3}.  (b) Time-domain waveforms of the microring control experiment with $-2.0\,\text{dBm}$ on-chip pump power, locking to 95$\%$ of the maximum photocurrent reading on the ADC. We directly measure a CAR of $42.0 \pm 0.7$ and a PGR of $38.6 \pm 0.8\,\text{kHz}$ in one of the two signal channels after stabilizing SFWM in the microring, before de-embedding chip coupling and detection system insertion losses.
    The first plot tracks the heater codes of the controlled microring (red) and an adjacent microring acting as a thermal aggressor (orange).
    The control heater DAC sweeps the microring from hot to cold for photocurrent calibration, then re-initializes it hot before locking it at the determined target ADC code.
    Both the CAR and the coincidence count rate are observed to peak at resonance during the calibration phase and then stabilize as the DAC locks to the target ADC code.
    We plot the coincidence rate and CAR for only one of the signal channels in this case to better show the trends during the calibration step, since points are dropped when the fitting algorithm fails to converge for low numbers of total counts in the histogram and this would occur much more frequently if the results from the two channels were summed.
    Once calibration finishes, the aggressor heater DAC pulses the adjacent microring as strongly as possible, switching between maximum and minimum heater codes, but the feedback loop maintains regulation with only slight variations observed in the ADC reading and pump transmission through the chip.
    On the other hand, when the feedback loop is disabled the same disturbance causes a snap-off of the microring resonance from the pump laser, ending photon-pair generation.}
    \label{fig4}
\end{figure}

To demonstrate the effectiveness of the feedback controller in stabilizing SFWM while other photonic devices on the same die are thermally tuned, we created a worst-case thermal disturbance by rapidly pulsing the heater of a nearby inactive photon-pair source microring from a different interleaved system site located 325\,\textmu m away. 
Fig. \ref{fig4} shows a diagram of the experiment and time-domain waveforms of the photon-pair statistics during the calibration sequence, locking phase, and thermal disturbances.
The adjacent microring steps between maximum and minimum heater settings, aggressively pushing the controlled microring's resonance closer to the point of instability by cooling it down.
However, the action of the control loop prevents the lock from being lost since it responds faster than the slow timescale of thermal crosstalk through the chip substrate.
With $-2.0\,\text{dBm}$ on-chip pump power the controller maintains a consistent off-chip pair rate of $38.6 \pm 0.8\,\text{kHz}$ and CAR of $42.0 \pm 0.7$ before and after the disturbance, with slight deviations during the thermal transients.
When the feedback is disabled and the heater DAC is set to a constant value, the same thermal disturbance causes a total loss of lock that cannot be restored without fully re-initializing the SFWM microring into the hot state and then cooling it back down again to meet the pump wavelength.


\subsubsection*{{\small Stabilized Photon-Pair Generation with Many Active Microrings}}

\begin{figure}[h!]
    \centering
    \includegraphics[width=\textwidth]{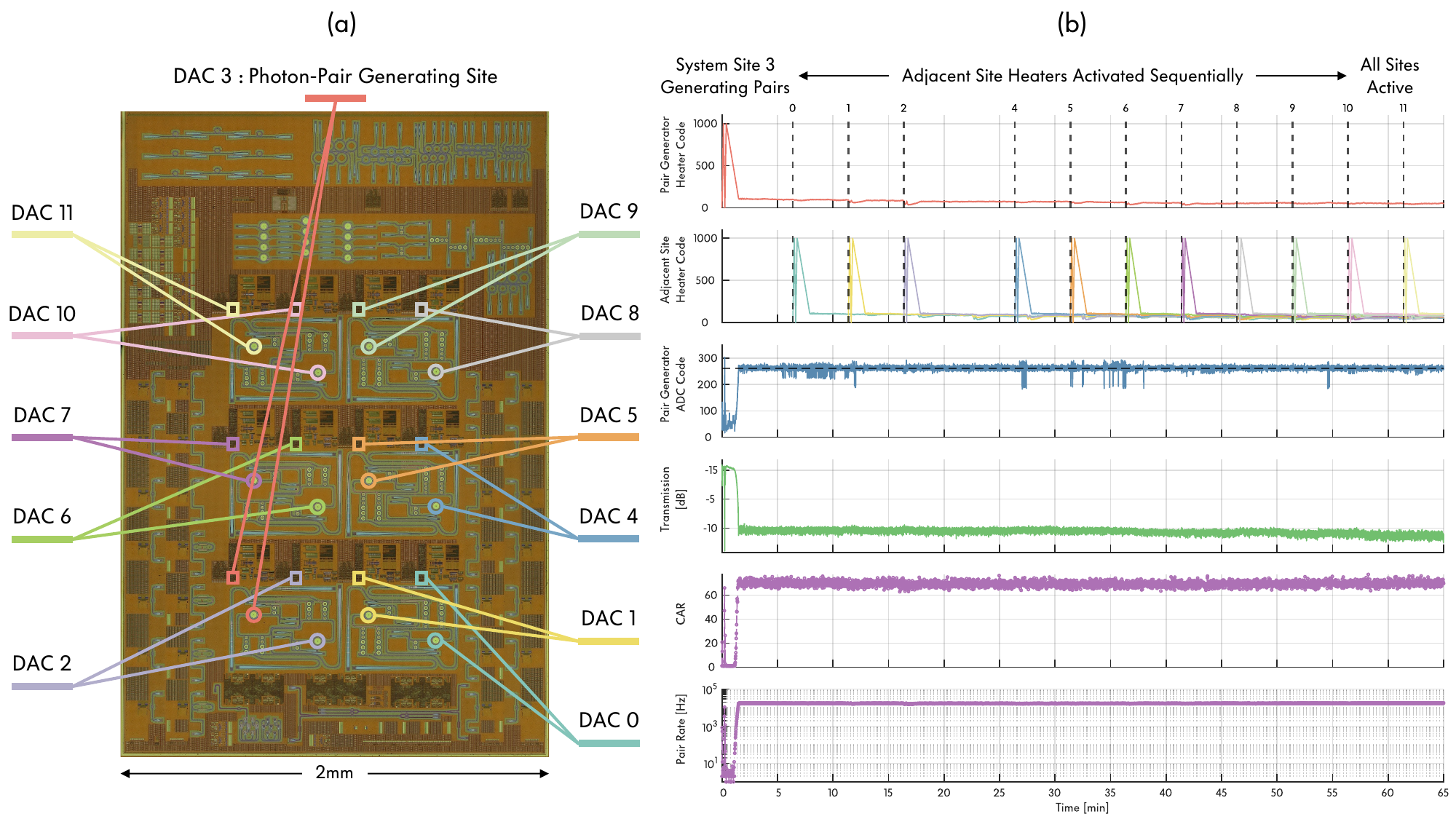}
    \caption{\textbf{Simultaneous operation of many photon-pair sources on a single chip:}  (a) Micrograph of the twelve SFWM microring control systems on the chip with each DAC and microring highlighted in the color of the corresponding DAC code plot. (b) Time-domain measurement of photon-pair generation in SFWM microring \#3 while the 11 adjacent sites are activated sequentially with delayed heater DAC waveforms over the course of an hour, emulating the operation of many photon-pair sources on the same chip. This experiment is comprised of $>3900$ photon-pair histograms. The off-chip detected PGR is $16.9 \pm 0.3\,\text{kHz}$ and CAR is $69.4 \pm 2.0$ with $-4.0\,$dBm on-chip pump power.}
    \label{fig5}
\end{figure}

For many microring-based photon-pair sources to operate together in an integrated system such as a multiplexed single-photon source \cite{collins2013integrated} or a quantum information processor, they must all lock to a common laser wavelength with each microring stabilized against thermal crosstalk from calibrating and locking the others.
Although we are limited to optically accessing only one on-chip photon source at a time by the placement of grating couplers and geometry of our fiber probes, we experimentally simulated the thermal environment of simultaneously locking all 12 SFWM microrings on our chip, depicted in Fig. \ref{fig5}(a), to test the control system under realistic operating conditions.
A microring near the center of the array was first locked to the pump laser using the previously described procedure. 
Then, time-delayed versions of the recorded DAC waveform were replayed on all 11 other SFWM microrings on the same chip over the course of approximately one hour.
This produces a similar thermal crosstalk profile to locking these sources -- in reality process variation will cause the final locked values of the heater DACs to vary among the microrings, but the largest and fastest thermal disturbances occur during the initial calibration phase. 
Fig. \ref{fig5}(b) shows the time-domain waveforms from this experiment.
With $-4.0\,\text{dBm}$ initial on-chip bus-waveguide pump power, the controlled SFWM microring produces a stable off-chip pair rate of $16.9 \pm 0.3\,\text{kHz}$ with a CAR of $69.4 \pm 2.0$ as the feedback controller compensates for thermal crosstalk from the other heaters.
Some variation in the pump laser transmission through the chip occurs due to drift in the alignment of the fiber probes, but the back-off of the regulation point from maximum photocurrent is sufficient to prevent a loss of lock (and such variations would not occur in a fully packaged die with a permanently glued fiber array, as discussed in Supplementary Information \S 5).
This experiment proves the ability of our integrated feedback control approach to enable the next generation of large scale quantum-photonic systems-on-chip to benefit from the advantages of microring-based photon sources.

\section*{Discussion}

In this work, we introduced the first electronic-photonic quantum system-on-chip (EPQSoC), which enables scalable control of microring resonator quantum photon-pair sources through the monolithic integration of silicon quantum photonics with complex control electronics on the same die.
Despite high-Q silicon microrings' extreme sensitivity to temperature variations, this system operates robustly without external stabilization from a thermoelectric cooler, greatly simplifying chip packaging.
The approach accounts for nonlinear effects present in such cavities and has permitted, to our knowledge, the highest Q-factor microring resonators controlled via integrated electronics to date.
Local feedback around each microring also avoids the complexity of thermal crosstalk cancellation schemes requiring detailed modelling and characterization of crosstalk among all devices on the same die \cite{milanizadeh2019canceling, gurses2022large}, which is yet to be demonstrated in nonlinear systems with microrings exhibiting self-heating or thermal bistability.
Although the on-chip circuits are already compact, the fast $\sim$60\,kHz sampling rate capability of the integrated photocurrent sensor means that the same circuit block could be time-multiplexed to control many microrings, requiring only the much smaller heater DAC and digital controller subcircuits to be duplicated (a detailed analysis of area scaling with multiplexed sensing is provided in Supplementary Information \S 3).
Thus, in future systems with tens to hundreds of controlled microrings the area overhead of the circuits can be much lower relative to the area of the photonics.
This is essential for maximizing the benefits of photonic integration as future silicon quantum-photonic systems approach the reticle size limit (typically 26\,mm\,$\times$\,33\,mm in modern CMOS tooling) and eventually the wafer size limit through tiling and reticle stitching \cite{seok2019wafer, jin2021seamless}.

Our use of a commercial CMOS foundry already geared towards high-yield volume production makes it possible to build large-scale quantum systems for which the small footprint and high performance of microring sources is crucial.
Furthermore, the recent introduction of a CMOS process (GlobalFoundries 45SPCLO \cite{rakowski2020cloprocess}) with native support for monolithic integration of photonics and electronics without requiring any post-processing or hacking of foundry design layers makes our control approach broadly accessible.
With its $\sim$2$\times$ thicker SOI layer improving the mode confinement in the silicon core and improved dielectric layer stack, our preliminary 45SPCLO device measurements in Supplementary Information \S 1 predict that optimized microring photon-pair sources in 45SPCLO could achieve $\sim$5.6\,GHz/mW$^2$ efficiency with the measured $\sim$$0.8 \times 10^6$ intrinsic Q-factors, a 1000$\times$ improvement over the results here.
Tighter mode confinement can also improve the CAR by reducing noise from spontaneous Raman scattering in the amorphous cladding materials.

An alternative way to scale up the control circuits is to use advanced packaging techniques such as die stacking, wafer bonding, or silicon interposers to bridge separate silicon photonics and control electronics chips together with a high density of electrical interconnects.
These approaches are already used in the microelectronics industry for high-end or high-volume commercial products \cite{lau2022recent, shekhar2024roadmapping, mahajan2019embedded}, but in a research and development setting they are costly and inaccessible compared to monolithic integration. 
With our approach, a whole quantum system can be designed in a single electronic design automation (EDA) tool flow, fabricated in a single tapeout, and packaged using standard flip-chip bonding to a printed circuit board.
This paradigm will enable rapid advances in the complexity of chip-scale quantum photonic systems.


\newpage
\section*{Methods}\label{sec_methods}

\subsection*{\small Chip Implementation}

Photonic device layouts were created using Berkeley Photonics Generator, an open-source Python tool for programmatic integrated photonics chip layout available at: \url{https://github.com/BerkeleyPhotonicsGenerator/BPG}.
Data preparation into process-specific mask design layers compliant with the CMOS foundry's geometric design rules (consisting of over 8000 checks) was then performed using Siemens Calibre software.
Custom high-density fill shapes were placed on the silicon and lower metal layers with $\sim 2.5\,\text{\textmu m}$ clearance around all photonic waveguides to meet the density requirements for CMP (chemical-mechanical polishing), while avoiding optical losses that would be caused by the default foundry density fill profile.
These shapes also reduce dishing effects originating in the low-density photonics regions that could otherwise impact the yield of nearby transistors.
The specific data preparation parameters we used are based on proprietary foundry information, which is available to customers under a non-disclosure agreement.
The analog and mixed-signal circuits within the photocurrent sensor were designed using Berkeley Analog Generator, another open-source Python tool for programmatic CMOS circuit design that wraps around Cadence Virtuoso, a standard commercial tool for integrated circuit design, to automate design and layout tasks.
It is available at \url{https://github.com/ucb-art/BAG_framework}.
Digital electronics design and top-level chip assembly were then performed using commercial digital-synthesis and place-and-route tools from Cadence to produce the final chip design.
Finally, physical verification of the full design was performed using Siemens Calibre DRC (design rule checking) and LVS (layout versus schematic comparison) software before sending the designs to the foundry.

\subsection*{\small Chip Fabrication}

The chips were fabricated as part of an MPW (multi-project wafer) run on GlobalFoundries' 45RFSOI process sponsored by Ayar Labs.
Some optional (non-critical) design rules were waived by the foundry, since the photonics layout contains geometric features that would reduce the performance or yield of transistors, for which the rule deck is optimized, but do not adversely affect photonic devices.
Nonetheless, transistors were fabricated in close proximity (within 10\,\textmu m) of photonics regions without any loss of functionality observed in over 20 chips tested.
Otherwise, the mask designs were treated in exactly the same way as standard electronics-only designs and processed using the ordinary manufacturing flow of the foundry.
The electrical pads on the wafers were plated with Cu pillars placed at a 250\,\textmu m pitch within the MPW chip area to enable electrical connections.

\subsection*{\small Electrical Packaging and Photonics Enablement}

Once diced, the chips were flip-chip soldered onto a 6-layer FR-4 HDI (high-density interconnect) PCB (printed circuit board) manufactured by Candor Industries.
The chips were underfilled with epoxy after soldering for improved mechanical stability.
To enable photonics functionality, the entire handle substrate of each packaged chip was removed in a single post-processing step using a XeF$_2$ dry etching tool (Xactix Xetch) in the Marvell Nanofabrication Laboratory at the University of California Berkeley, an approach we previously developed for integrating classical photonic links into CMOS microprocessors \cite{sun2015nature, stojanovic2018monolithic}.
The thin ($<200\,\text{nm}$) buried oxide (BOX) layer in the 45RFSOI CMOS process would otherwise lead to excessive guided mode tunnelling leakage into the Si substrate --- a problem not encountered in conventional silicon photonics wafers with 2--3\,\textmu m thick BOX.
This layer, on which the silicon layer with transistors and photonic devices sits, acts as an etch stop that prevents them from being removed from the back side along with the silicon handle.
In this process the performance of electronics was not affected by the removal of the silicon handle, circumventing the need to mask off the circuits and selectively etch only the photonics regions.
Lithographic masking to contain the etch in this way does, however, improve heatsinking of the circuits and mechanical reliability of the chip, and has been implemented at wafer scale for volume production of photonic link chiplets in the same CMOS process \cite{wade2020teraphy, wade2021teraphy}.


\subsection*{\small Electro-Optical Testing}

Light was coupled in and out of the packaged and etched chips using tapered and lensed fibers from OZ Optics with 5\,\textmu m spot size matching the mode-field diameter of the VGCs.
The fibers were held on custom-machined holders attached to 3-axis Thorlabs NanoMax positioners and manually aligned to the VGCs under a microscope with a long working distance metallurgical objective (10X Mitutoyo Plan Apo), selected to allow sufficient clearance for the fiber holders.
A Keysight 81608A C-band tunable laser source with a continuous sweep synchronized to a Agilent 81635A InGaAs optical power sensor was used for optical transmission characterization of the photonics and initial bring-up and testing of the electronic circuits.
The Q-factors and FSRs reported in Fig. \ref{fig2} and Supplementary Information \S 1 were measured by locally fitting Lorentzian curves to the signal, pump, and idler resonances with several varying on-chip powers of roughly -20\,dBm to -50\,dBm to provide reasonable estimates of uncertainty while avoiding distortion from self-heating at higher laser powers.

Since the tuning range of the microrings is less than their free spectral range, arbitrary pump wavelengths cannot be supported.
Instead, a pump laser wavelength is chosen around 0.25\,nm to 0.5\,nm longer than the measured SFWM microring resonance wavelength after performing a transmission sweep.
The low tuning range of the SFWM microrings was caused by a correctable design bug in the heater layout code that caused the entire microring to be filled with a disk-shaped heater instead of a narrower track near the waveguide, reducing the thermal tuning efficiency.
Corrections to this issue that have been implemented on subsequent chip tapeouts are described in Supplementary Information \S 2.

\subsection*{\small Feedback Controller Implementation}

In the experiments reported here, we read out and control the on-chip ADC and DAC through a serial scan chain interface from an external field-programmable gate array (FPGA), bypassing on-chip digital feedback control circuits to allow full customization of the tuning algorithm.
The FPGA relays commands to a PC through a USB interface, allowing arbitrary control algorithms and command sequences to be executed using Python code. The controller we implemented can later be hardened to custom logic in future chips, and given the simplicity of our 3-step finite state machine controller it can easily fit in the same area as the existing circuit (which is similar in nature, but cannot lock as close to the maximum photocurrent due to some hardcoded parameters).
The long latency of the FPGA to Python software interface limits the control loop update rate to $\sim$10\,Hz when running from Python code --- this could be improved to $\sim$2\,kHz by implementing the same algorithms on the FPGA directly (limited by the 10\,MHz scan clock and 4,674-bit length of the scan chain), or $\sim$60\,kHz with fully custom logic on the chip operating at the ADC sample rate (including a $512 \times$ on-chip averaging filter we used here to reduce noise).

\subsection*{\small Single-Photon Measurements}

A Pure Photonics PPCL200 low-noise Micro-Integrable Tunable Laser Assembly (\textmu ITLA) was used as the pump in the single-photon counting experiments.
It was routed through a 3-paddle manual fiber polarization controller (FPC) to a set of fiber-coupled bandpass (channel add-drop) thin-film filters with $>100\,\text{dB}$ extinction in aggregate at the signal and idler wavelengths in order to filter out the noise from ASE (amplified spontaneous emission) in the laser and spontaneous Raman scattering in the fiber connecting the laser to the input of this filter (Raman noise generated after this filter is not attenuated).
The input laser power to the chip was monitored through the 10\% tap of a 90:10 power splitter between the filter and the lensed fiber probe coupled to the input VGC.
Estimated on-chip pump powers are reported by accounting for the insertion loss of the input VGC.
Light transmitted through the chip was collected by a second lensed fiber coupled to an output VGC and sent to an AC Photonics 1$\times$2 DWDM (dense wavelength division multiplexing) channel-dropping filter with a 100\,GHz-wide passband centered at 1545.3\,nm (ITU grid channel C40) to extract the signal photons onto its ``pass'' port.
The signals photons are detected on a pair of SNSPDs after a 50:50 beamsplitter to enable $g^{(2)}$ measurements.
A second identical filter was cascaded in the signal photon path to provide a total of $>120\,\text{dB}$ isolation of the residual pump light.
The idler and residual pump photons exiting the ``reflect'' port of the first DWDM filter were connected to a different pair of cascaded DWDMs with a 100\,GHz-wide passband centered at 1559.8\,nm (ITU grid channel C22).
These filters route the idler photons to their ``pass'' port (also with $>120\,\text{dB}$ of isolation of the residual pump) and route the residual pump to the reflect port, which is connected to a Thorlabs S154C photodiode to measure the pump laser transmission through the chip, which is used to estimate the VGC insertion losses.
The input FPC is adjusted to maximize the pump transmission through the chip before aligning the SFWM microring, since the VGCs efficiently couple a single polarization of light.
With the SFWM microring aligned, signal and idler photons are detected using three channels of a Quantum Opus One SNSPD system, which operates the single-photon detectors in a 2.4\,K cryostat.
This system also requires manual FPC adjustments on each channel to maximize the photon detection efficiency (PDE).
The PDEs of the idler and two signal SNSPD channels, including fiber and FPC losses, were characterized using variably attenuated laser light to be 63\%, 77\%, and 74\%, respectively.
The various manual FPC adjustments could be eliminated in a packaged system with polarization maintaining fibers.

\subsection*{\small Coincidence Count Analysis}

Photon detection events from the SNSPDs were recorded using a Swabian Instruments Time Tagger 20 TCSPC system that collected timestamps of trigger events with 1\,ps time-bin quantization.
The timestamps were then processed using the ``start-stop'' measurement software provided with the instrument to produce a histogram of time-of-arrival differences between signal and idler photons.
The timing jitters of the signal and idler SNSPDs were characterized by the manufacturer to be $\sim$80\,ps and the timing jitter of the time tagger is 35\,ps.
We observed histograms with coincidence peaks having much wider widths in time, which we attribute primarily to the time uncertainty corresponding to the linewidth (photon energy uncertainty) of the SFWM process in the microring --- see \cite{yard2022chip} for a detailed discussion of how the timing uncertainty of single-photon detection interacts with cavity linewidth.
To extract the PGR and CAR from a given histogram, we fitted an offset Gaussian (a good approximation of the convolution of the three different sources of timing uncertainty) within $\pm$0.8\,ns of the coincidence peak.
We then used a 320\,ps wide coincidence window (roughly $\pm 2\,\sigma$, kept constant throughout all measurements) to extract the total numbers of coincidences and accidentals, and compute their ratio (the CAR) \cite{bienfang2023single}. The PGR is then defined as the ratio of the coincidences to the integration time of the histogram, and the PGR and corresponding CAR can be traded off by varying the choice of coincidence window (although there is no single correct choice, we have picked a standard one here).

To verify the quantum nature of our source and characterize multi-pair emissions, we measure the conditional second-order correlation function $g^{(2)}(0)$ using a standard three-detector setup \cite{beck2007comparing}.
We first perform correlation measurements between the idler and two signal channels to determine their relative delays.
After digitally compensating for those delays, we use the same 320\,ps coincidence window as for measuring the CAR and PGR.
We then compute the time-integrated $g^{(2)}(0)$ using the following expression: $g^{(2)}(0) = \nicefrac{N_\text{I,S1,S2} N_\text{I} }{N_\text{I,S1} N_\text{I,S2}}$,
where $N_\text{I,S1,S2}$ is the number of threefold coincidences between the signal channels and the idler, $N_\text{I}$ is the number of single counts in the idler detector and $N_\text{I,S1}$ and $N_\text{I,S2}$ are the numbers of twofold coincidences between the idler and the individual signal channels.


\subsection*{\small Potential for On-Chip Pump Filtering and SNSPD Integration}

The experiments reported here relied on an external pump ASE cleanup filter at the input side and external pump rejection and signal-idler demultiplexing filters on the output side to isolate the photon pairs generated in the SFWM microring from noise and residual pump light.
They were then detected by off-chip SNSPDs in a cryostat separate from the rest of the system operating at room temperature.
Ultimately large-scale quantum systems will need to include these functions on a single chip to fully realize the benefits of an integrated platform with single-photon sources and detectors.
Supplementary Information \S 4 reviews and discusses possible routes for integrating SNSPDs and prospects for operating CMOS circuits and thermally-tuned microring resonators at the requisite few-K cryogenic temperatures for SNSPD compatibility.
Furthermore, we previously showed that high-order cascaded microring filters located on the same chip as an SFWM source can achieve the large ($>$95\,dB) levels of stopband rejection required for these applications \cite{gentry2018monolithic}.
Based on measured microring Q-factors, these types of filters could also achieve state of the art ($<$1\,dB) passband insertion losses if actively aligned to compensate for process variations.
We included such filters (with $4^\text{th}$ and $6^\text{th}$-order cascaded microring variants) along with tuning circuits on this chip.

\begin{figure}[h!]
    \centering
    \includegraphics[width=\textwidth]{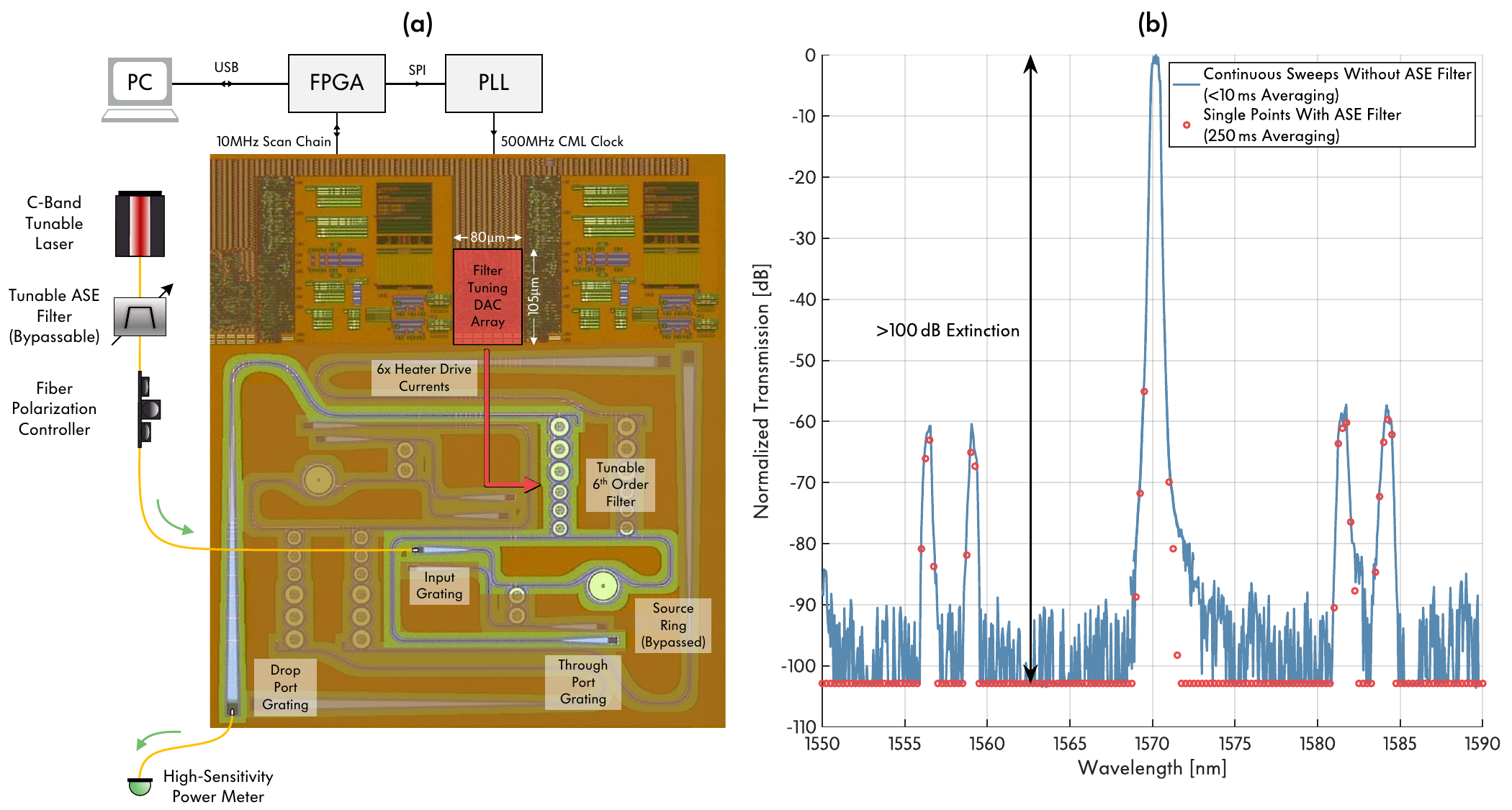}
    \caption{\textbf{(Extended Data) High-Extinction Pump Rejection Filter:}  (a) Measurement setup for characterizing the extinction ratio of the integrated pump rejection filters. The filters are aligned by using an EDFA in place of the tunable laser to intentionally generate broadband ASE noise, and then maximizing the drop port power meter readings using a Nelder-Mead optimization algorithm. The tunable laser is then swapped back in to take a transmission spectrum.  (b) Normalized transmission spectrum of input to drop port response of the aligned filter using a series of stitched continuous sweeps with varying power meter range settings (blue curve) showing $\sim$95\,dB extinction that is limited by ASE noise of the laser source passing through the aligned passband. When an external ASE filter is enabled and the integration time is increased, the noise floor falls below the minimum detectable level of the power meter, indicating $>$100\,dB extinction at an offset wavelength of one FSR of the photon-pair source, which is required for fully on-chip pump filtering (red points). The filter uses a Vernier scheme to extend the FSR, resulting in less extinction ($\sim$60\,dB) at the FSRs of the cascaded sub-filters, which are selected not to align with either the pump laser or other single-photon channel when the pump rejection filter is aligned to a particular signal or idler output wavelength. Limited heater tuning range prevented the on-chip filters from being aligned to a photon-pair source resonance wavelength on all tested chips, which is correctable by improving the source and filter heater designs and slightly adjusting the microring radii to target specific resonance wavelengths, reducing the initial error between the source and filter resonances. We have characterized the insertion loss of this filter to be $<$1\,dB using measurements at a separate test site with through port and drop port grating couplers oriented in the same direction, allowing the same fiber to measure both ports one at a time \cite{cabanillas2022monolithically}.}
    \label{fig6}
\end{figure}

Fig. \ref{fig6} (extended data) shows a measurement of the $>$100\,dB contrast between the passband and stopband consistent needed for on-chip filtering, consistent with our previous results in this platform.
However, correctable design bugs limited the tuning range of these filters, while also nominally placing the signal and idler passbands at the same wavelength, instead of nominally placing them two free spectral ranges of the SFWM microring apart.
This prevented them from being aligned for such a system demonstration, and they were bypassed.
We also bypassed an included $2^\text{nd}$-order cascaded microring filter with 40\,dB of stopband rejection intended for pump cleanup because we found that the large ASE noise of the available pump laser source (Pure Photonics PPCL200) required a much higher degree of stopband rejection for a coincidence peak to be observable above the noise floor.
In the future, a higher-order filter can be used for pump cleanup and the initial (unaligned) wavelengths of the signal and idler separation filters can be placed closer to the relevant resonance wavelengths of the SFWM microring to enable alignment of a fully integrated photon-pair generation system without external filtering.

\backmatter

\bmhead{Supplementary information}

An accompanying supplementary information file is provided.

\bmhead{Data availability statement}

Data sets generated during the current study are available from the corresponding authors on reasonable request.



\newpage

\newpage

\newpage

\newpage

\newpage

\newpage

\newpage
\bibliography{references_main}

\end{document}


\title{Supplementary Information: Scalable Feedback Stabilization of Quantum Light Sources on a CMOS Chip}




\author*[1]{Danielius~Kramnik}\email{kramnik@berkeley.edu}
\author*[2]{Imbert~Wang}\email{imbert@bu.edu}
\author*[3]{Anirudh~Ramesh}\email{anirudh@u.northwestern.edu}
\author[2]{Josep~M.~Fargas~Cabanillas}
\author[2]{\DJ{}or\dj{}e~Gluhovi\'{c}}
\author[1]{Sidney~Buchbinder}
\author[1]{Panagiotis~Zarkos}
\author[1]{Christos~Adamopoulos}
\author[3]{Prem~Kumar}
\author[1]{Vladimir~M.~Stojanovi\'{c}}
\author[2]{Milo\v{s}~A.~Popovi\'{c}}

\affil[1]{\orgdiv{Department of Electrical Engineering and Computer Sciences}, \orgname{University of California Berkeley}, \orgaddress{\city{Berkeley}, \state{CA}  \postcode{94709}, \country{USA}}}

\affil[2]{\orgdiv{Department of Electrical and Computer Engineering and Photonics Center}, \orgname{Boston University}, \orgaddress{\city{Boston}, \state{MA}  \postcode{02215}, \country{USA}}}

\affil[3]{\orgdiv{Center for Photonic Communication and Computing, Department of Electrical and Computer Engineering}, \orgname{Northwestern University}, \orgaddress{\city{Evanston}, \state{IL}  \postcode{60208}, \country{USA}}}

\maketitle


\newpage
\section{Spontaneous Four-Wave Mixing Microring Design}

\subsection{Pair Generation Rate and Optimal Coupling Condition}

\begin{figure}[h]
    \includegraphics[width=\textwidth]{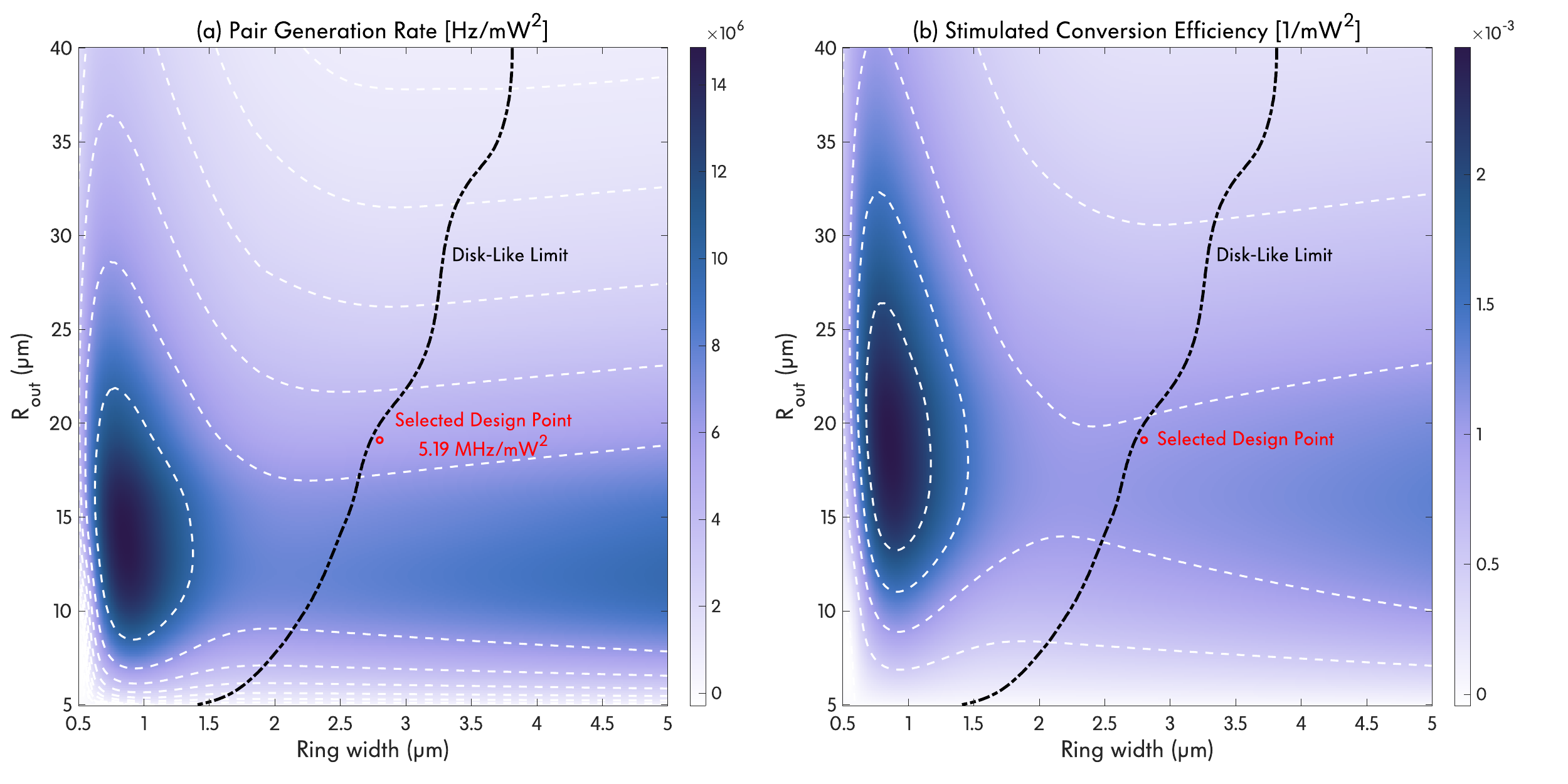}
    \caption{Simulated (a) correlated photon-pair generation rate and (b) stimulated four-wave mixing efficiency at 1\,mW pump power versus microring outer radius and waveguide width in GlobalFoundries' 45RFSOI CMOS process.}
    \label{fig_pgr_stim_design_trends}
\end{figure}

The photon-pair source ring was designed with the goal of optimizing the spontaneous four-wave mixing (SFWM) pair generation rate (PGR) while accommodating the \textit{p}- and \textit{n}-doped radial spokes used for carrier sweepout and photodetection.
Fig \ref{fig_pgr_stim_design_trends}(a) shows the expected PGR per pump power squared for circular microrings as a function of radius and width for the silicon thickness and cladding layers used in GlobalFoundries’ 45nm SOI CMOS process.
This theoretical PGR (before accounting for nonlinear and carrier-induced loss mechanisms, which reduce the PGR at high pump powers) is given by the following equation \cite{gentry2018scalable}:
\begin{equation}\label{eqn_pgr}
    I_{\text{pair}}
    =
    \omega_\text{p}^2 \beta_\text{FWM}^2
    \left(
    \frac{ 2 r_\text{p,e} }{ r_\text{p,tot}^{2} }
    \right)^2
    \frac{ 2 r_\text{i,e} r_\text{s,e} }{ r_\text{i,tot} r_\text{s,tot} }
    \frac{ r_\text{s,tot} + r_\text{i,tot} }{ \left(2 \pi \Delta \nu_\text{FSR} \right)^2 + \left( r_\text{s,tot} + r_\text{i,tot} \right)^2 }
    P_\text{p}^2
\end{equation}

\noindent where $r_\text{p/s/i,e}$ are the coupled-mode theory (CMT) extrinsic decay rates of the pump/signal/idler energy amplitudes due to coupling to the external bus waveguide, $r_\text{p/s/i,tot}$ are the CMT decay rates of the pump/signal/idler energy amplitudes due to all mechanisms, $P_\text{p}$ is the input pump power, $\Delta \nu_\text{FSR}$ is the free spectral range (FSR) mismatch caused by dispersion between the signal, pump, and idler modes, and $\beta_\text{FWM}$ is the nonlinear four-wave mixing coefficient.
Note that the CMT coupling rates can be expressed in terms of Q-factor as $r = \omega / 2 Q$, allowing experimentally measured FSR mismatch and resonance Q-factors to predict the PGR efficiency if $\beta_\text{FWM}$ is known.
For a given microring geometry, $\beta_\text{FWM}$ and $\Delta \nu_\text{FSR}$ can be determined numerically using a 2D electromagnetic mode solver and the extrinsic decay rates can be designed by adjusting the ring-bus coupling geometry. The procedure for calculating $\beta_\text{FWM}$ is described in the next section, and $\Delta \nu_\text{FSR}$ is calculated using the procedure described in \cite{gentry2014discrete}.

In our design we expect the ring-bus coupling and intrinsic loss rates to have little variation across $\pm1$ FSR of the cavity, so we assume $r_\text{p,e} = r_\text{s,e} = r_\text{i,e} = r_\text{e}$, $r_\text{p,o} = r_\text{s,o} = r_\text{i,o} = r_\text{o}$, and $r_\text{p,tot} = r_\text{s,tot} = r_\text{i,tot} = r_\text{tot}$. The relationship between $r_\text{e}$ and $r_\text{o}$ (with $r_\text{tot} = r_\text{e} + r_\text{o}$) that maximizes the PGR is then found by setting the derivative of $I_\text{pair}$ with respect to coupling \cite{gentry2018scalable} equal to zero, yielding:
\begin{equation}
    r_\text{e} = \frac{ 4 }{ 3 } r_\text{o}
    \label{eqn_optimal_pgr_coupling}
\end{equation}

\subsection{Calculation of the Nonlinear Coefficient}

In a photonic waveguide, the nonlinear coefficient depends on the third-order susceptibility tensor of the material $\overline{\overline{\boldsymbol{\chi}}}^{(3)}$ and an overlap integral between the interacting mode fields.
For the signal mode, it is given by \cite{zeng2014design}:
\begin{equation}\label{eqn_beta_fwm}
    \beta_\text{FWM,s}
    =
    \frac{
    \frac{ 3 }{ 16 } \varepsilon_\text{o}
    \int \mathrm{d}^3 \mathbf{x}
    \left(
    \mathbf{E}_\text{s}^* \cdot \overline{\overline{\boldsymbol{\chi}}}^{(3)} : \mathbf{E}_\text{i}^* \mathbf{E}_\text{p}^2
    \right)
    }{
    \sqrt{
    \int \mathrm{d}^3 \mathbf{x}
    \left( \frac{1}{2} \varepsilon \left| \mathbf{E}_\text{s} \right|^2 \right)
    }
    \sqrt{
    \int \mathrm{d}^3 \mathbf{x}
    \left( \frac{1}{2} \varepsilon \left| \mathbf{E}_\text{i} \right|^2 \right)
    }
    \int \mathrm{d}^3 \mathbf{x}
    \left( \frac{1}{2} \varepsilon \left| \mathbf{E}_\text{p} \right|^2 \right)
    }
    \equiv
    \frac{ 3 \chi^{(3)}_{1111} }{ 4 \varepsilon_\text{o} n^4_\text{Si} V_\text{eff} }
\end{equation}

\noindent where $\mathbf{E}_\text{p/s/i}$ are the electric field vectors of the pump/signal/idler modes within the silicon core, $n_\text{Si}$ is the refractive index of silicon, and $c$ is the speed of light.
The third-order nonlinearity of silicon is much stronger than any of the cladding materials, so we assume $\overline{\overline{\boldsymbol{\chi}}}^{(3)} = 0$ in the cladding and restrict all volumes of integration to the silicon core.
The second expression defines $V_\text{eff}$, the effective nonlinear interaction volume corresponding to the volume of bulk nonlinear medium in which uniform fields with the same energy would have equal nonlinearity.
With full permutation symmetry of $\overline{\overline{\boldsymbol{\chi}}}^{(3)}$, $\beta_\text{FWM,s} = \beta_\text{FWM,i} = \beta_\text{FWM,p}^* \equiv \beta_\text{FWM}$ (the Manley-Rowe relations \cite{boyd2008nonlinear} --- note that another common convention uses $2\beta_\text{FWM,s} = 2\beta_\text{FWM,i} = \beta_\text{FWM,p}^*$ \cite{ramirez2011degenerate}, with the factor of $2$ being accounted for in the CMT equations and ultimately yielding the same results).
Thus, adjusting the device dimensions to minimize $V_\text{eff}$ produces a stronger nonlinear coefficient.

To evaluate the numerator integral in $\beta_\text{FWM}$, first consider the nonzero elements of $\overline{\overline{\boldsymbol{\chi}}}^{(3)}$. Silicon has a diamond cubic crystal structure with m3m crystallographic point-group symmetry, meaning that $\overline{\overline{\boldsymbol{\chi}}}^{(3)}$ has 4 independent elements, and each element of the tensor can be written as:
\begin{equation}
    \chi^{(3)}_{ijkl}
    =
    \left(
    \chi^{(3)}_{1111} - \chi^{(3)}_{1122} - \chi^{(3)}_{1212} - \chi^{(3)}_{1221}
    \right)
    \delta_{ijkl}
    +
    \chi^{(3)}_{1122}
    \delta_{ij} \delta_{kl}
    +
    \chi^{(3)}_{1212}
    \delta_{ik} \delta_{jl}
    +
    \chi^{(3)}_{1221}
    \delta_{il} \delta_{kj}
    \label{eq_general}
\end{equation}
where the indices $i, j, k, l \in \{ x, y, z \}$, and $\hat{x}$, $\hat{y}$, and $\hat{z}$ denote the [100], [010], and [001] crystal lattice directions, respectively. Since the signal and idler wavelength have small offsets relative to the pump wavelength and all three are well below the bandgap, we can safely assume Kleinman symmetry \cite{boyd2008nonlinear} (i.e. $\chi^{(3)}_{1122} \approx \chi^{(3)}_{1221} \approx \chi^{(3)}_{1212}$), reducing the equation to:
\begin{equation}
    \chi^{(3)}_{ijkl}
    =
    \left(
    \chi^{(3)}_{1111} - 3 \chi^{(3)}_{1122}
    \right)
    \delta_{ijkl}
    +
    \chi^{(3)}_{1122}
    \left(
    \delta_{ij} \delta_{kl} + \delta_{ik} \delta_{jl} + \delta_{il} \delta_{kj}
    \right)
\end{equation}

The nonlinear coefficient can then to be written in the following form:
\begin{equation}
    \beta_\text{FWM}
    =
    \frac{
    \frac{ 3 }{ 16 } \varepsilon_\text{o}
    \left(
    \chi^{(3)}_{1111} A + \chi^{(3)}_{1122} B
    \right)
    }{
    \sqrt{
    \int_\text{Si} \mathrm{d}^3 \mathbf{x}
    \left( \frac{1}{2} \varepsilon \left| \mathbf{E}_\text{s} \right|^2 \right)
    }
    \sqrt{
    \int_\text{Si} \mathrm{d}^3 \mathbf{x}
    \left( \frac{1}{2} \varepsilon \left| \mathbf{E}_\text{i} \right|^2 \right)
    }
    \int_\text{Si} \mathrm{d}^3 \mathbf{x}
    \left( \frac{1}{2} \varepsilon \left| \mathbf{E}_\text{p} \right|^2 \right)
    }
\end{equation}
with:
\begin{equation}
    A
    =
    \int_\text{Si} \mathrm{d}^3 \mathbf{x}
    \left(
    E_{\text{s},x}^* E_{\text{i},x}^* E_{\text{p},x}^2 + E_{\text{s},y}^* E_{\text{i},y}^* E_{\text{p},y}^2 + E_{\text{s},z}^* E_{\text{i},z}^* E_{\text{p},z}^2
    \right)
    \approx
    \int_\text{Si} \mathrm{d}^3 \mathbf{x}
    \left(
    \left| E_x \right|^4 + \left| E_y \right|^4 + \left| E_z \right|^4
    \right)
\end{equation}

\begin{equation}
\begin{alignedat}{2}
    B
    &=
    \int_\text{Si} \mathrm{d}^3 \mathbf{x} \,
    \Big(
    &&E_{\text{s},x}^* E_{\text{i},x}^* E_{\text{p},y}^2
    + E_{\text{s},x}^* E_{\text{i},x}^* E_{\text{p},z}^2
    + E_{\text{s},y}^* E_{\text{i},y}^* E_{\text{p},x}^2
    \\
    & &&+ E_{\text{s},y}^* E_{\text{i},y}^* E_{\text{p},z}^2
    + E_{\text{s},z}^* E_{\text{i},z}^* E_{\text{p},x}^2
    + E_{\text{s},z}^* E_{\text{i},z}^* E_{\text{p},y}^2
    \\
    & &&+ 2 E_{\text{s},x}^* E_{\text{i},y}^* E_{\text{p},x} E_{\text{p},y}
    + 2 E_{\text{s},x}^* E_{\text{i},z}^* E_{\text{p},x} E_{\text{p},z}
    + 2 E_{\text{s},y}^* E_{\text{i},x}^* E_{\text{p},y} E_{\text{p},x}
    \\
    & &&+ 2 E_{\text{s},y}^* E_{\text{i},z}^* E_{\text{p},y} E_{\text{p},z}
    + 2 E_{\text{s},z}^* E_{\text{i},x}^* E_{\text{p},z} E_{\text{p},x}
    + 2 E_{\text{s},z}^* E_{\text{i},y}^* E_{\text{p},z} E_{\text{p},y}
    \Big)
    \\
    &\approx
    \int_\text{Si} \mathrm{d}^3 \mathbf{x} \,
    \Big(
    &&E_{x}^{*2} ( E_y^2 + E_z^2 )
    + E_{y}^{*2} ( E_x^2 + E_z^2 )
    + E_{z}^{*2} ( E_x^2 + E_y^2 )
    \\
    & &&+ 4 | E_x |^2 | E_y |^2
    + 4 | E_x |^2 | E_z |^2
    + 4 | E_y |^2 | E_z |^2
    \Big)
\end{alignedat}
\end{equation}
\noindent where the approximations assume that the pump, signal, and idler mode solutions are identical, which introduces small inaccuracy across the relatively narrow wavelength span of the calculation ($\pm1$\,FSR of the resonator, which is typically $\sim 10\,\text{nm}$).


\begin{figure}[b]
\includegraphics[width=\textwidth]{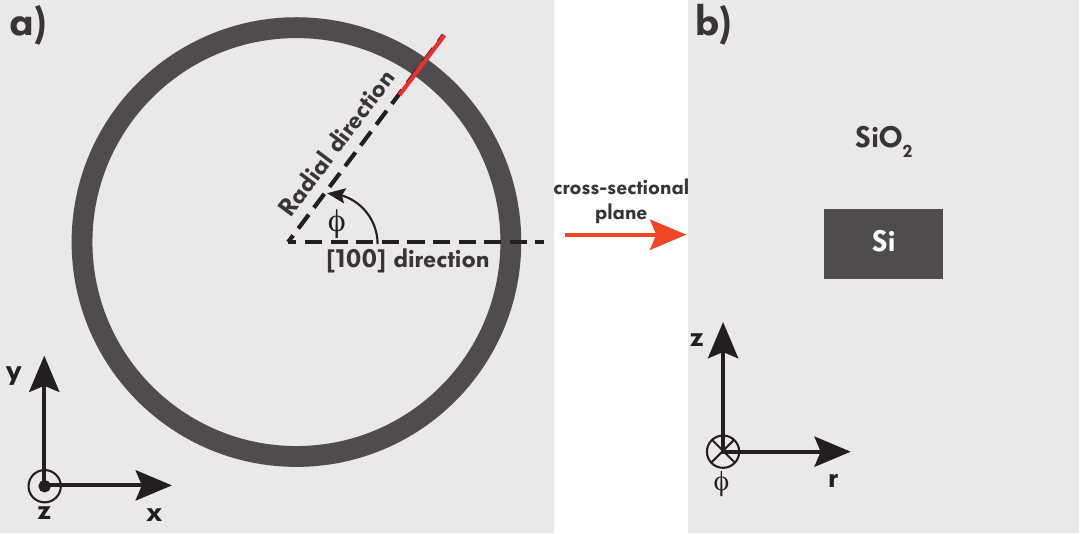}
\caption{Coordinate system used to compute the average overlap integral of the optical mode with the \(\chi^{(3)}\) nonlinearity across the microring resonator. (a) shows the top view of the ring waveguide. (b) shows the cross-sectional view of the microring waveguide.}
\label{chi_3_average}
\end{figure}

Given the circular symmetry of ring resonators, we use a bent waveguide mode solver to obtain the electric field in cylindrical coordinates $( E_r, E_\phi, E_z )$ for a cross-section of the microring, rather than $( E_x, E_y, E_z )$ within the whole device.
The expressions for the $A$ and $B$ coefficients are then transformed from rectangular to cylindrical coordinates, projecting them onto the $[100]$-oriented silicon crystal of the SOI wafer, which is done with the following vector field coordinate transform:
\begin{equation}
    \begin{pmatrix}
        E_r \\
        E_\phi \\
        E_z \\
    \end{pmatrix}
    =
    \begin{pmatrix}
        \cos \phi    & \sin \phi   & 0 \\
        -\sin \phi    & \cos \phi    & 0 \\
        0            & 0            & 1 \\
    \end{pmatrix}
    \begin{pmatrix}
        E_x \\
        E_y \\
        E_z \\
    \end{pmatrix}
\end{equation}
\noindent where $\phi$ is the azimuthal angle between $\hat{r}$ and the [100] silicon crystal lattice vector as shown in Fig. \ref{chi_3_average}.
Averaging over $\phi$ and simplifying the final expressions, we obtain:
\begin{equation}
    A
    =
    2 \pi
    \iint_\text{Si} 
    r \, \mathrm{d}r \, \mathrm{d}z
    \left(
    |E_z|^4 + \frac{3}{4} |E_r|^4 + \frac{3}{4} |E_\phi|^4 + \frac{1}{4} E_r^2 E_\phi^2 + \frac{1}{4} E_r^{*2} E_\phi^{*2} + |E_r|^2 |E_\phi|^2
    \right)
\end{equation}

\begin{equation}
\begin{split}
    B
    =
    2 \pi
    \iint_\text{Si} 
    r \, \mathrm{d}r \, \mathrm{d}z
    \left(
    E_z^{*2} \left( E_r^2 + E_\phi^2 \right)
    + E_z^2 \left( E_r^{*2} + E_\phi^{*2} \right)
    + \frac{3}{4} |E_r|^4 + \frac{3}{4} |E_\phi|^4
    \right.
    \\
    \left.
    + \frac{1}{4} E_r^{*2 }E_\phi^2 + \frac{1}{4} E_\phi^{*2}E_r^2 + 4 |E_z|^2 \left(|E_r|^2 + |E_\phi|^2 \right) +|E_r|^2|E_\phi|^2
    \right)
\end{split}
\end{equation}
\noindent Note that the denominator of $\beta_\text{FWM}$ makes the overall expression independent of the choice of normalization in the mode simulation. These equations allow $\beta_\text{FWM}$ and $V_\text{eff}$ to be calculated numerically for varying microring dimensions, which is in turn used to calculate $I_\text{pair}$.

\subsection{Microring Geometry Design Procedure}

The microring radius and width were selected with the goal of maximizing $I_\text{pair}$ while keeping the heavily doped radial spokes away from the fundamental waveguide mode, preventing large optical losses from free carrier absorption (FCA) and scattering from the waveguide contacts.
In a bent waveguide, the energy of the fundamental optical mode concentrates closer to outer sidewall of the waveguide as the bend becomes sharper.
We define a ``disk-like'' mode as one that is pushed so much closer to the outer sidewall than the inner sidewall that the SFWM nonlinear interaction is insensitive with respect to perturbations in the inner sidewall radius ($\mathrm{d}V_\text{eff} / \mathrm{d}r_\text{in} \approx 0$).
In other words, the fraction of the mode power at the inner sidewall is so small that the mode is minimally affected by any index perturbation in that region.
This makes it possible to introduce doped contacts with a large imaginary (lossy) refractive index component without adding significant optical loss to the fundamental mode.
The boundary between disk-like and ordinary ring-like modes is plotted over the predicted $I_\text{pair}$ in Fig. \ref{fig_pgr_stim_design_trends}, with disk-like modes being to the right of the line --- the boundary is approximate since the dependence on inner radius approaches zero asymptotically.
The disk-like limit can be reached by decreasing the microring radius or increasing the waveguide width.
To calculate the intrinsic decay rate as a function of radius and width, which sets all decay rate parameters, we assumed a base $4.6\,\text{dB/cm}$ of absorption and scattering loss based on straight waveguide measurements at $1550\,\text{nm}$ in the same process \cite{orcutt2012open} and added bending losses from the mode simulations.

A local maximum occurs in $I_\text{pair}$ in the ring-like region, and a plateau in $I_\text{pair}$ occurs in the disk-like region at a waveguide width corresponding to a completely filled disk, but neither is at a suitable design point for a contacted ring. 
The peak in the ring-like region is stronger, but introducing contacts would result in excessive amounts of optical loss.
On the other hand, the plateau in the disk-like region occurs with extremely wide waveguide widths, creating a different set of issues.
First, the doped contacts would be far from the optical mode, making the electric field applied from the \textit{p-i-n} junctions very weak in the photogeneration region and hindering carrier extraction.
Second, the small inner radius of the contacted ring would leave little space for the heater in the center of the ring required for thermal tuning.
Instead, as a compromise between these requirements and the goal of maximizing $I_\text{pair}$, we selected an area in the design space near the boundary of the disk-like region that maximizes the four-wave mixing strength.
We picked two different outer radii ($18.8\,\text{\textmu m}$ and $19.1\,\text{\textmu m}$) and waveguide widths ($2.0\,\text{\textmu m}$ and $2.8\,\text{\textmu m}$) across the 12 different SFWM microrings in the systems we taped out, also varying the geometry and number of doped contact spokes.
Once the geometrical parameters of the microring were selected, a wrapped pulley coupler with the target $r_\text{e}$ and propagation constant matched to the fundamental mode was designed using FDTD simulations.
We followed a standard procedure used previously for designing wrapped couplers in high-speed microring modulators with similar geometry in the same CMOS process \cite{hosseini2010systematic,shainline2013depletion}.

The experiments reported in the main text all use a design with $r_\text{out} = 19.1\,\text{\textmu m}$ and $w = 2.8\,\text{\textmu m}$, which we found to be the best combination.
At the time of the design, we used the classical simulated four-wave mixing efficiency, plotted in Fig. \ref{fig_pgr_stim_design_trends}(b), rather than the PGR as the figure of merit.
The equation for this efficiency, also derived in \cite{gentry2018scalable}, is slightly different, leading to a different optimum point in the design:
\begin{equation}\label{eqn_eta_stim}
    \eta_{\text{stim}}
    =
    \omega_\text{p}^2 \beta_\text{FWM}^2
    \left(
    \frac{ 2 r_\text{p,e} }{ r_\text{p,tot}^{2} }
    \right)^2
    \frac{ 2 r_\text{s,e} }{ r_\text{s,tot}^2 }
    \frac{ 2 r_\text{i,e} }{ \left(2 \pi \Delta \nu_\text{FSR} \right)^2 + r_\text{i,tot}^2 }
    P_\text{p}^2
\end{equation}
By optimizing the expression for $I_\text{pair}$ rather than $\eta_{\text{stim}}$ in future designs, we expect a 50\,\% increase in the PGR to be possible in future designs (our model predicts $5.2\,\text{MHz}/\text{mW}^2$ at the selected design point versus $7.8\,\text{MHz}/\text{mW}^2$ at the optimal PGR point).




\subsection{Impact of Microring Wavelength, FSR, and Q-Factor Variability}

Variation in the Q-factors and FSRs of microring photon-pair sources can affect the rate and wavelength with which the output photons are produced.
In previous literature, inhomogeneity in the FSRs of different photon-pair sources fabricated in silicon photonics, which lead to differences in output photon wavelengths, has not come up as an issue when demonstrating quantum interference between microring-based sources \cite{silverstone2015qubit, faruque2018chip, lu2020threedim, llewellyn2020teleportationchip, arrazola2021xanadu}.
These experiments, however, used microrings fabricated nearby on the same chip.
Photon-pair sources fabricated on different chips, either from different locations on the same wafer or different wafer lots entirely, could exhibit enough variability in waveguide thickness or etching to make the mismatch of FSRs non-negligible and reduce the visibility of their quantum interference effects.
This concern will need to be addressed further down the line as these systems scale beyond the point where they can be contained on a single chip.
Fabricating photon-pair sources in a CMOS foundry capable of producing systems-on-chip with deep submicron transistors is appealing because the tight process control needed to yield massive numbers of transistors should translate to lower variability in microring geometry.
To characterize this, we measured the resonance wavelengths, FSRs, and Q-factors of the photon-pair source microring at 2 different locations of 6 separate packaged dies available from the same multi-project wafer (MPW) run (the 12 system sites on each chip are used for design splits of the microring, but those 12 designs are also copied on a separate device test area located around 2\,mm away from the system sites).
The full results are reported in Table \ref{table:source_ring_variability}.

\begin{table}[t!]
\centering

\tiny

\begin{subtable}[c]{\textwidth}
\centering
\begin{tabular}{ccccccccccc}
\multicolumn{1}{c}{\textit{Die}}
&\multicolumn{1}{c}{\textit{Site Type}}
&\multicolumn{1}{c}{$\lambda_\text{s}$ [nm]}
&\multicolumn{1}{c}{$\lambda_\text{p}$ [nm]}
&\multicolumn{1}{c}{$\lambda_\text{i}$ [nm]}
&\multicolumn{1}{c}{$Q_\text{o,s}$}
&\multicolumn{1}{c}{$Q_\text{e,s}$}
&\multicolumn{1}{c}{$Q_\text{o,p}$}
&\multicolumn{1}{c}{$Q_\text{e,p}$}
&\multicolumn{1}{c}{$Q_\text{o,i}$}
&\multicolumn{1}{c}{$Q_\text{e,i}$}
\\
\hline
\textbf{B9} & \textbf{System} & \textbf{1545.29} & \textbf{1552.50} & \textbf{1559.79} & \textbf{64.4\,k} & \textbf{57.7\,k} & \textbf{113.1\,k} & \textbf{68.2\,k} & \textbf{102.8\,k} & \textbf{62.7\,k} \\
B9 & Test & 1545.11 & 1552.33 & 1559.62 & 98.0\,k & 74.9\,k & 72.4\,k & 59.4\,k & 124.9\,k & 65.2\,k \\
C2 & System & 1544.82 & 1552.02 & 1559.32 & 95.8\,k & 80.4\,k & 74.6\,k & 62.9\,k & 103.9\,k & 58.3\,k \\
C2 & Test & 1544.27 & 1551.49 & 1558.78 & 84.5\,k & 73.0\,k & 100.3\,k & 68.4\,k & 81.8\,k & 57.9\,k \\
C4 & System & 1547.92 & 1555.14 & 1562.44 & 106.7\,k & 76.7\,k & 108.2\,k & 75.5\,k & 120.0\,k & 67.8\,k \\
C4 & Test & 1547.82 & 1555.04 & 1562.35 & 97.8\,k & 73.9\,k & 125.9\,k & 67.6\,k & 132.1\,k & 66.7\,k \\
C5 & System & 1543.19 & 1550.40 & 1557.69 & 92.9\,k & 75.6\,k & 106.2\,k & 72.1\,k & 108.2\,k & 69.9\,k \\
C5 & Test & 1543.53 & 1550.74 & 1558.03 & 92.9\,k & 73.9\,k & 99.6\,k & 66.2\,k & 97.5\,k & 54.7\,k \\
C6 & System & 1548.02 & 1555.24 & 1562.55 & 77.2\,k & 71.1\,k & 106.4\,k & 77.5\,k & 103.7\,k & 62.1\,k \\
C6 & Test & 1546.59 & 1553.81 & 1561.11 & 78.6\,k & 71.6\,k & 117.4\,k & 82.6\,k & 97.6\,k & 58.7\,k \\
C7 & System & 1547.26 & 1554.48 & 1561.77 & 68.6\,k & 63.0\,k & 93.4\,k & 66.8\,k & 136.2\,k & 71.6\,k \\
C7 & Test & 1546.73 & 1553.95 & 1561.25 & 98.9\,k & 80.0\,k & 121.4\,k & 80.4\,k & 85.9\,k & 55.4\,k \\
\hline
\multicolumn{2}{c}{Means:}
& $1545.88$
& $1553.10$
& $1560.39$
& $88.0$\,k
& $72.6$\,k
& $103.2$\,k
& $70.6$\,k
& $107.9$\,k
& $62.6$\,k
\\
\multicolumn{2}{c}{Std. devs.:}
& 1.73
& 1.74
& 1.74
& 13.2
& 6.5
& 16.7
& 7.1
& 17.2
& 5.7
\\
\hline
\multicolumn{2}{c}{Best intra-die:}
& 0.10
& 0.09
& 0.09
& $<0.1$\,k
& 0.6\,k
& 6.6\,k
& 5.1\,k
& 6.1\,k
& 0.4\,k
\\
\multicolumn{2}{c}{Worst intra-die:}
& 1.43
& 1.44
& 1.44
& 33.7\,k
& 17.0\,k
& 40.7\,k
& 13.6\,k
& 50.3\,k
& 16.2\,k
\\
\multicolumn{2}{c}{Best die-die:}
& 0.10
& 0.11
& 0.10
& 0.3\,k
& $<0.1$\,k
& 0.2\,k
& 0.2\,k
& $<0.1$\,k
& 0.4\,k
\\
\multicolumn{2}{c}{Worst die-die:}
& 4.83
& 4.84
& 4.86
& 42.4\,k
& 22.7\,k
& 53.5\,k
& 23.1\,k
& 54.4\,k
& 16.9\,k
\\
\end{tabular}
\caption{Variability data for resonance wavelength and fitted Q-factors. Uncertainties in the reported Q-factors for each microring are all on the order of 1\,k, exact values are omitted for clarity.}
\end{subtable}

\vspace{1em}

\begin{subtable}[c]{\textwidth}
\centering
\begin{tabular}{ccccccccc}
\multicolumn{1}{c}{\textit{Die}}
&\multicolumn{1}{c}{\textit{Site Type}}
&\multicolumn{1}{c}{$\Delta \nu_\text{s}$ [GHz]}
&\multicolumn{1}{c}{$\Delta \nu_\text{p}$ [GHz]}
&\multicolumn{1}{c}{$\Delta \nu_\text{i}$ [GHz]}
&\multicolumn{1}{c}{FSR$_\text{sp}$ [GHz]}
&\multicolumn{1}{c}{FSR$_\text{pi}$ [GHz]}
&\multicolumn{1}{c}{$\Delta \nu_\text{FSR}$ [GHz]}
&\multicolumn{1}{c}{PGR [MHz/mW$^2$]}
\\
\hline 
\textbf{B9} & \textbf{System} & $\mathbf{6.38 \pm 0.14}$ & $\mathbf{4.54 \pm 0.05}$ & $\mathbf{4.94 \pm 0.05}$ & $\mathbf{901.49 \pm 0.38}$ & $\mathbf{902.92 \pm 0.08}$ & $\mathbf{1.42 \pm 0.43}$ & $\mathbf{3.29 \pm 0.13}$ \\ 
B9 & Test & $4.58 \pm 0.04$ & $5.92 \pm 0.14$ & $4.49 \pm 0.12$ & $901.71 \pm 0.35$ & $903.24 \pm 0.33$ & $1.52 \pm 0.44$ & $3.16 \pm 0.15$ \\
C2 & System & $4.44 \pm 0.02$ & $5.66 \pm 0.10$ & $5.15 \pm 0.05$ & $900.91 \pm 0.37$ & $903.44 \pm 0.13$ & $2.53 \pm 0.29$ & $3.03 \pm 0.10$ \\
C2 & Test & $4.96 \pm 0.10$ & $4.75 \pm 0.09$ & $5.68 \pm 0.11$ & $902.25 \pm 0.04$ & $903.82 \pm 0.09$ & $1.57 \pm 0.09$ & $3.50 \pm 0.11$ \\
C4 & System & $4.34 \pm 0.10$ & $4.34 \pm 0.03$ & $4.43 \pm 0.04$ & $898.70 \pm 0.12$ & $901.56 \pm 0.09$ & $2.86 \pm 0.08$ & $5.61 \pm 0.11$ \\
C4 & Test & $4.62 \pm 0.33$ & $4.43 \pm 0.56$ & $4.34 \pm 0.27$ & $899.35 \pm 1.99$ & $901.62 \pm 2.03$ & $3.72 \pm 1.75$ & $7.21 \pm 0.15$ \\
C5 & System & $4.66 \pm 0.05$ & $4.51 \pm 0.12$ & $4.53 \pm 0.07$ & $903.27 \pm 0.88$ & $904.72 \pm 0.98$ & $1.45 \pm 0.75$ & $4.77 \pm 0.19$ \\
C5 & Test & $4.72 \pm 0.15$ & $4.86 \pm 0.03$ & $5.49 \pm 0.07$ & $902.50 \pm 0.04$ & $904.95 \pm 0.24$ & $2.46 \pm 0.27$ & $4.25 \pm 0.26$ \\
C6 & System & $5.23 \pm 0.08$ & $4.30 \pm 0.06$ & $4.94 \pm 0.12$ & $899.20 \pm 0.52$ & $901.13 \pm 0.42$ & $1.84 \pm 0.50$ & $3.75 \pm 0.07$ \\
C6 & Test & $5.18 \pm 0.16$ & $3.98 \pm 0.14$ & $5.24 \pm 0.02$ & $899.96 \pm 0.23$ & $902.44 \pm 0.31$ & $2.47 \pm 0.09$ & $4.19 \pm 0.12$ \\
C7 & System & $5.90 \pm 0.07$ & $4.95 \pm 0.04$ & $4.09 \pm 0.03$ & $898.98 \pm 0.03$ & $901.28 \pm 0.07$ & $2.30 \pm 0.06$ & $3.16 \pm 0.06$ \\
C7 & Test & $4.38 \pm 0.03$ & $3.99 \pm 0.06$ & $5.70 \pm 0.03$ & $899.99 \pm 0.11$ & $901.69 \pm 0.10$ & $1.70 \pm 0.17$ & $5.34 \pm 0.08$ \\
\hline
\multicolumn{2}{c}{Means:}
& $4.95 \pm 0.63$
& $4.69 \pm 0.60$
& $4.92 \pm 0.54$
& $900.70 \pm 1.53$
& $902.73 \pm 1.33$
& $2.16 \pm 0.70$
& $4.27 \pm 1.27$
\\
\hline
\multicolumn{2}{c}{Best intra-die:}
& $0.06 \pm 0.17$
& $0.10 \pm 0.56$
& $0.09 \pm 0.27$
& $0.22 \pm 0.52$
& $0.06 \pm 2.03$
& $0.10 \pm 0.62$
& $0.13 \pm 0.20$
\\
\multicolumn{2}{c}{Worst intra-die:}
& $1.80 \pm 0.15$
& $1.38 \pm 0.14$
& $1.61 \pm 0.05$
& $1.34 \pm 0.37$
& $1.31 \pm 0.53$
& $1.00 \pm 0.80$
& $2.18 \pm 0.10$
\\
\multicolumn{2}{c}{Best die-die:}
& $0.04 \pm 0.34$
& $0.01 \pm 0.15$
& $0.01 \pm 0.13$
& $0.03 \pm 0.26$
& $0.08 \pm 2.04$
& $0.02 \pm 0.29$
& $0.00 \pm 0.16$
\\
\multicolumn{2}{c}{Worst die-die:}
& $2.04 \pm 0.17$
& $1.93 \pm 0.20$
& $1.58 \pm 0.12$
& $4.57 \pm 0.88$
& $3.83 \pm 0.49$
& $2.30 \pm 1.80$
& $4.18 \pm 1.54$
\\
\end{tabular}
\caption{Variability data for resonance FWHM linewidth ($\Delta \nu_\text{s,p,i}$), free spectral range (FSR), FSR mismatch ($\Delta \nu_\text{FSR}$), and predicted PGR using Eq. \ref{eqn_pgr}. The FWHM linewidths are related to the resonance Q-factors via $Q_\text{tot} = \lambda_\text{res} / \Delta \lambda_\text{FWHM}$ where $Q_\text{tot} = ( Q_\text{o}^{-1} + Q_\text{e}^{-1} )^{-1}$ and $\Delta \nu_\text{FWHM} = \Delta \lambda_\text{FWHM} \cdot c / \lambda_\text{res}^2$.}
\end{subtable}

\caption{Summary of variability between 6 fully packaged dies. Multiple wavelength sweeps from 1540\,nm to 1560\,nm are taken with varying laser power to estimate the uncertainties (reported as standard deviation) in the measured quantities. The laser power is kept below $-20$\,dBm to avoid FSR measurement errors arising from differing amounts of self-heating of each resonance during the scan. The system site used for the experiments reported in the main text is highlighted in bold (die label B9).}
\label{table:source_ring_variability}
\end{table}

In general, we find that Q-factors and predicted pair generation rates have an appreciable degree of variability with $>$2$\times$ difference between the best and worst microrings in terms of predicted PGR.
However, the intra-die FSR variation is small compared to the linewidths of the resonances, even in the worst case, meaning that sources on the same chip should exhibit a high visibility of quantum interference (e.g. Hong-Ou Mandel effect) once they are locked to a common pump wavelength.
From die to die, we find that FSRs can be matched very closely with binning, but that the worst-case differences are on the order of a resonance linewidth, which means there would be an appreciable degree of mismatch between the wavelengths of heralded photons when locked to the same laser, and a reduction of visibility.
We do, however, note that the way in which we performed the XeF$_2$ etch step may be exaggerating the intra-die variations because these chips were processed in separate batches over the course of $>$1 year, sometimes with different numbers of total dies in the process chamber.
Since the BOX layer is very thin, the amount of over-etching of the BOX can slightly influence the waveguide mode.
This issue is avoided in GlobalFoundries' next-generation CMOS photonics process, 45SPCLO \cite{rakowski2020cloprocess}, with a thicker BOX layer that eliminates the need for the etching step.
The spread in resonance wavelengths between dies is greater than the tuning range of the heater driver, but this is mainly caused by a design bug that is addressed later in Supplementary Information \S 2.3, with measurements in an improved heater and microring design showing sufficient range to cover the fractional FSR variability measured here.
It would be valuable to carry out future variability studies with wafer-level testing equipment and uniform post-processing, if any, to obtain better statistics and insights on how to bin the dies by proximity and location on the wafer.

To avoid binning in future systems, design techniques for reducing FSR and resonance wavelength variation can be implemented.
As one example, adiabatic microrings have been shown to exhibit less variability in resonance wavelength and FSR due to process errors than uniform microrings \cite{su2014reduced, mirza2024experimental}.
The waveguide widths of different sections of a non-uniform microring can also be specifically selected to cancel out the group-index derivative with respect to the width, substantially reducing FSR variability \cite{ouyang2019silicon}.
Our microring optimization technique assumes uniform waveguide width -- future work is needed to extend this to non-uniform microrings so that PGR and FSR variability can be simultaneously considered and optimized.
Finally, active compensation schemes such as coupling to an auxiliary microring cavity can be used to trim the FSRs of different microrings to match \cite{gentry2014tunable}, but this adds substantial control complexity and increases $V_\text{eff}$, so developing passive compensation methods is preferable.
The ring-bus couplings ($Q_\text{e}$) can also be actively controlled via interferometric coupling to optimize the PGR across fabrication variations \cite{wu2022optimization}.



\subsection{Roadmap for Microring Performance Improvements in CMOS}

The waveguide loss of 4.6\,dB/cm at 1550\,nm wavelength used in our device optimization limits the achievable intrinsic Q-factor.
This loss value was measured in 45RFSOI for a 670\,nm-wide single-mode waveguide \cite{orcutt2012open} and includes both material absorption and sidewall scattering losses.
We used this available loss measurement as a conservative estimate in our design procedure, but would expect wider undoped multi-mode waveguides to have lower sidewall scattering losses.
The fact that our resonators have comparable average intrinsic Q-factors (90\,k to 100\,k) to that predicted with this conservative estimate (around 110\,k) suggests that there is room for improvement in the design of our spoked contacts and \textit{p-i-n} junction dopings, which can introduce additional scattering and free-carrier absorption losses.
This issue is exacerbated by the fact that we used heavily-doped and silicided contact dopings for these junctions, which introduce more free carrier losses and dark current than lighter dopings.
This can be addressed in the future by including the complex index of refraction of the dopings in the waveguide mode simulations, moving them further away from the optical field, and using graded dopings with lower doping density away from the contact vias.

The data in \cite{orcutt2012open} show an absorption peak around 1520\,nm wavelength that we attribute to the silicon nitride stress liner ($\approx 100\,$nm thick) that is deposited across the whole chip \cite{yang2004dual} and present over the silicon waveguides (see extended data Fig. 1 in \cite{sun2015nature} for an SEM image, available \href{https://www.nature.com/articles/nature16454/figures/5}{here}).
This absorption peak occurs due to Si-H and N-H bonds in the SiH$_4$ and NH$_3$ precursors used in low-temperature PECVD deposition of the nitride films \cite{frigg2019low}, which do not use a deposition recipe optimized for optical loss because the 45RFSOI CMOS process is intended for electronics use.
This loss peak can be avoided in 45RFSOI by switching wavelengths from C-band to O-band, which boosted the intrinsic Q-factor of microrings by more than a factor of $2 \times$ in the previous measurements \cite{orcutt2012open}.
The improved waveguide confinement at shorter wavelengths can also boost the $\beta_\text{FWM}$ nonlinear coefficient and reduce spontaneous Raman scattering noise generated in the amorphous cladding materials.
Thus, we expect the CAR to be improved in O-band both from increased pair generation rate and reduced noise level.

\begin{figure}[t!]
    \centering
    \includegraphics[width=\textwidth]{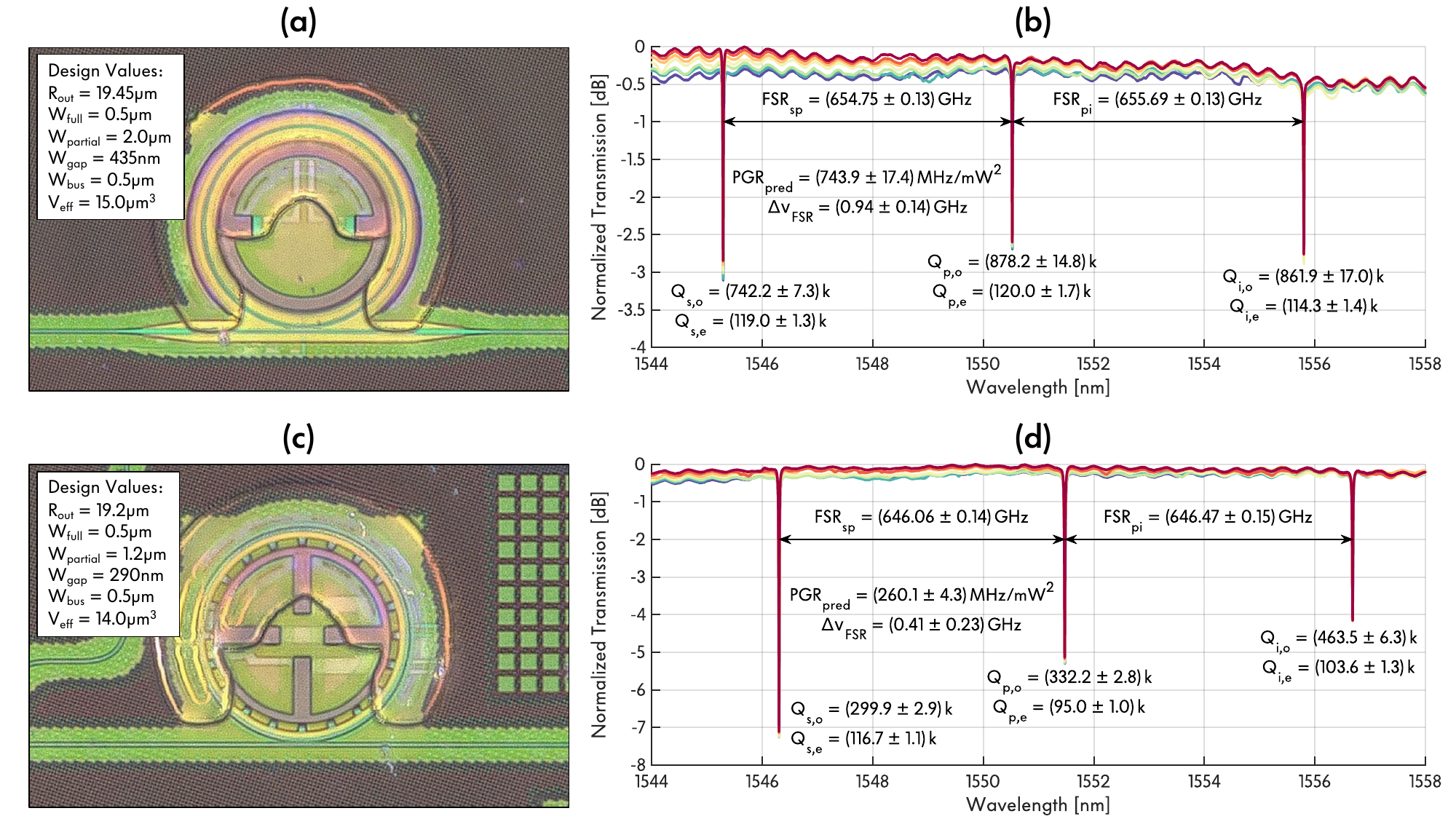}
    \caption{(a) Micrograph of full-rib SFWM-optimized microring in GlobalFoundries' new 45SPCLO CMOS photonics process. (b) Normalized transmission spectra and performance metrics of full rib microring. (c) Micrograph of half-rib SFWM-optimized microring in 45SPCLO \cite{rakowski2020cloprocess}. (b) Normalized transmission spectra and performance metrics of half-rib microring. Note that in both cases there is a small gap between the heater and \mbox{\textit{p-i-n}} junction contacts to electrically isolate the two, but it is difficult to discern in the micrographs. Transmission spectra are taken with varying on-chip laser powers ranging from around $-25\,\text{dBm}$ to $-55\,\text{dBm}$ to provide reasonable estimates of uncertainty while avoiding distortion from self-heating at higher laser powers.}
    \label{fig_clo_q_factors}
\end{figure}

Beyond these improvements in the 45RFSOI platform, we can achieve even greater gains in the performance of these photon-pair sources using GlobalFoundries' new 45SPCLO CMOS photonics process \cite{rakowski2020cloprocess}, which has specific optimizations for photonic device performance while maintaining compatibility with densely integrated electronics.
The photonics are fabricated using a thicker silicon layer than the transistors ($\sim$160\,nm vs. sub-100\,nm) that does not have a nitride stress liner and achieves \mbox{$<2\,\text{dB/cm}$} propagation loss in both C-band and O-band single-mode waveguides (exact performance metrics and a photonic layer stackup with geometric and optical parameters are available under nondisclosure agreement with the foundry). 
The photonic waveguides can also be partially etched, allowing rib waveguides to be fabricated.
This makes it easier to sweep out photogenerated carriers and reduces sidewall scattering losses.

Fig. \ref{fig_clo_q_factors} shows micrographs and preliminary measurements of two SFWM-optimized microrings that we designed and fabricated in 45SPCLO.
The first variant is a rib waveguide device with $V_\text{eff} = 15.0\,\text{\textmu m}^3$ that achieves intrinsic Q-factors of $\sim$750-850\,k and FSR mismatch of $\sim$0.9\,GHz, resulting in a predicted PGR efficiency of $\sim$740\,MHz/mW$^2$ when loaded with an extrinsic (ring-bus coupling) Q-factor of $\sim$120\,k.
The second variant is a half-rib waveguide device with $V_\text{eff} = 14.0\,\text{\textmu m}^3$ that achieves intrinsic Q-factors of $\sim$300-400\,k and FSR mismatch of $\sim$0.4\,GHz, resulting in a predicted PGR efficiency of $\sim$260\,MHz/mW$^2$ when loaded with an extrinsic Q-factor of $\sim$100\,k.
In both cases, the predicted PGR efficiency is limited by overcoupling to the bus waveguide beyond the optimum given in Eq. \ref{eqn_optimal_pgr_coupling} and could be improved by widening the coupling gap.
The predicted PGR efficiency of the rib waveguide microring can be increased to $\sim$5.6\,GHz/mW$^2$ by scaling-up its $Q_\text{e}$-factors by $5.3 \times$ while the predicted PGR efficiency of the half-rib waveguide microring can be increased to $\sim$530\,MHz/mW$^2$ by scaling-up its $Q_\text{e}$-factors by $2.6 \times$.
These values represent improvements of $1000\times$ and $100\times$ over our results in the 45RFSOI platform, respectively.
The half-rib waveguide design is appealing because it achieves lower $V_\text{eff}$ and FSR mismatch, but it used closer spacing of doped contacts to the waveguide core than the rub waveguide design, resulting in lower intrinsic Q-factors.
This suggests there is room for further work to optimize these designs and achieve even better performance in 45SPCLO.


\section{On-Chip Control Circuit Design and Calibration}

\subsection{Analog Frontend Circuit Design and Calibration}

\begin{figure}[t!]
    \centering
    \includegraphics[width=\textwidth]{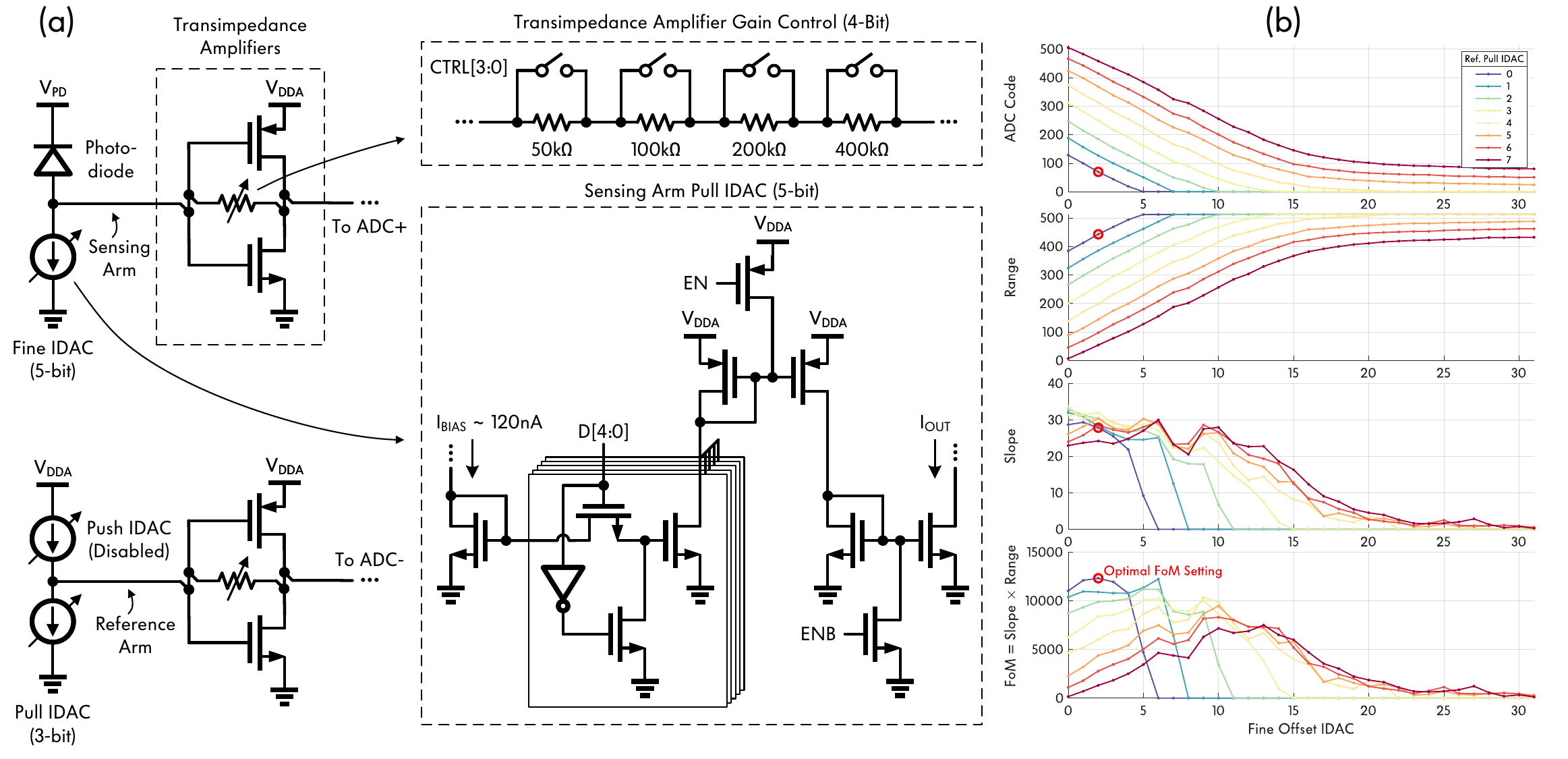}
    \caption{(a) Analog frontend circuits with pseudo-differential TIAs for transducing the photocurrent and driving a differential SAR ADC, with insets showing the programmable gain and current-mode DACs (IDACs). (b) Example AFE calibration scans taken with 200\,k$\Omega$ gain setting to establish the circuit operating point before scanning the microring resonance.}
    \label{fig_afe_cal}
\end{figure}

Fig. \ref{fig_afe_cal}(a) shows a detailed schematic of the analog frontend (AFE) circuitry in the on-chip photocurrent sensor shown in the main text.
Photocurrent from the SFWM microring is first amplified by a pseudo-differential pair of transimpedance amplifiers (TIAs) with digitally programmable gain varying from $50\,\text{k}\Omega$ to $750\,\text{k}\Omega$.
Gain adjustment is implemented using four binary-weighted resistors per TIA ranging from $50\,\text{k}\Omega$ to $400\,\text{k}\Omega$ with transmission gates connected in parallel, allowing each to either be switched in to the TIA feedback path or bypassed.
Current-mode digital to analog converter (IDAC) circuits allow additional offset currents to be pushed or pulled from the two TIA inputs, which are used to adjust the operating point of the analog frontend circuits to optimize their dynamic range and compensate for dark current and offsets created by PVT (process, voltage, and temperature) variations affecting the device parameters.
The amplified photocurrent is then digitized by a 9-bit analog to digital converter (ADC) implemented using a self-timed successive approximation register (SAR) topology \cite{liu2020time}.
This SAR ADC is clocked through a $16 \times$ frequency divider from the chip's high-speed clock, resulting in a 31.25\,MHz sample rate with the 500\,MHz high-speed clock frequency used in our experiments.
Its output is then averaged 512 times using an on-chip digital accumulator to reduce noise, producing a $\sim$61\,kHz update rate post-averaging.

To take advantage of the full input range of the ADC in spite of PVT and dark current induced offsets, we sweep the current DACs connected to the two TIA inputs to find the setting that maximizes a figure of merit defined by the product of the range and photocurrent response of the TIAs.
This step is performed at the start of each experiment before the microring is aligned to the pump laser so that photocurrent does not disrupt the calibration --- it needs to be redone every time the gain setting is changed through the feedback resistor configuration.
The 9-bit ADC output code ranges from 0 to $2^{9}-1 = 511$, with increasing photocurrent producing a larger code.
Therefore, the difference between the maximum code (511) and the initial setting before any photocurrent is induced in the SFWM microring defines the range available to the photocurrent sensor.
We use the current DAC connected to the SFWM microring TIA input as a proxy for photocurrent in order to find the response of the sensor, which we define as the magnitude of the change in ADC code per LSB (least-significant bit) step of the current DAC.
This response is maximized when the TIAs and ADC are biased midscale because the low loop gain of the compact inverter-based TIAs produces soft clipping distortion when the output voltage approaches the power supply rails.
Midscale initial biasing is undesirable, however, because it discards half of the available range and puts the locked photocurrent level too close to the maximum ADC code.
It is preferable to place the initial ADC reading closer to minimum code to increase the range and soft-clip at low photocurrents rather than high photocurrents so that deviations from the lock point can more easily be resolved by the control loop.
Fig. \ref{fig_afe_cal}(b) shows an example of such a calibration sweep and the chosen operating point that  maximizes the figure of merit, taken with the TIA gain set to 200\,k$\Omega$.
Note that the pull-down IDAC and sensing arm photocurrent apply currents to the TIA with opposite polarity, so the IDAC lowers the ADC code to increase the range and the photocurrent then undoes the action of the IDAC, pulling the ADC code back up.
For scaling-up the control of integrated photonic systems it is important to automate such calibration procedures rather than fine-tuning the offset DACs and other circuit settings on the bench before each experiment.


\subsection{Heater DAC Design and Effective Wavelength Resolution}

\begin{figure}[t!]
    \centering
    \includegraphics[width=\textwidth]{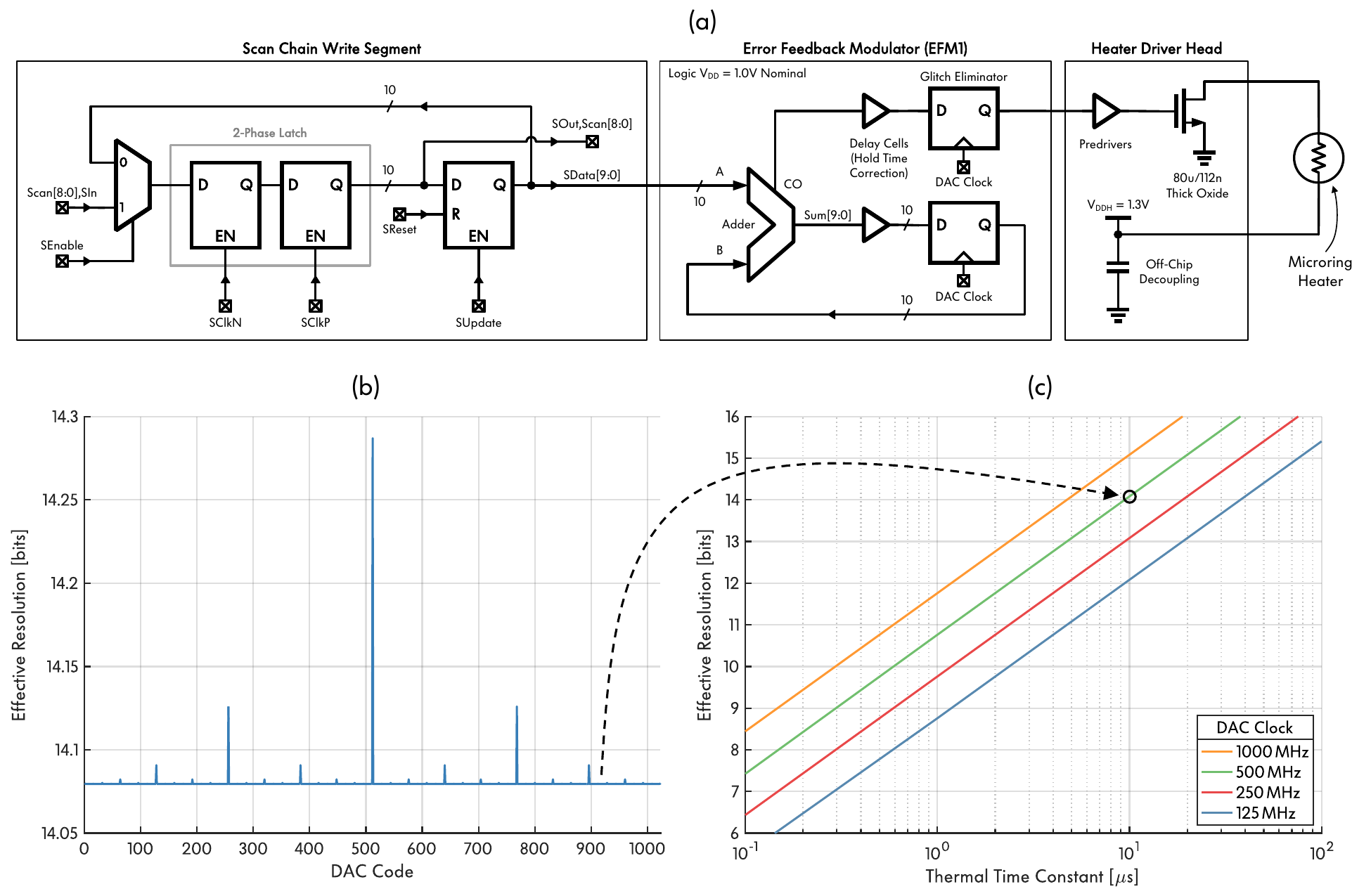}
    \caption{(a) Delta-sigma heater DAC schematic (actual design carried out using digital synthesis and place and route tools from Verilog code). (b) Simulated effective resolution due to residual ripple versus DAC code for $f_\text{clk} = 500\,\text{MHz}$ and $\tau_\text{th} = 10\,\text{\textmu s}$. (c) Simulated effective resolution trends for varying $f_\text{clk}$ and $\tau_\text{th}$.}
    \label{fig_dac_sim}
\end{figure}

The heater driver DAC uses a switching output stage driven by a delta-sigma modulator to quickly switch the SFWM microring heater on and off at the high-speed clock rate, relying on the slow thermal time constant of the heater and microring to filter out the switching pattern to a steady value \cite{sun2013pdm}.
Fig. \ref{fig_dac_sim}(a) shows a schematic of this heater driver circuit.
The physical number of bits is determined by the width of the accumulator register in the delta-sigma modulator, allowing the mean (DC) value of the heater power to be set with arbitrary precision.
However, the residual ripple around this DC value is set by the ratio of the high-speed clock to the thermal time constant ($\tau_\text{th}$) of the heater, meaning that beyond a certain point adding bits to the accumulator register does not improve the effective resolution with which the control loop can tune the microring, while still increasing area and power consumption, and reducing timing margin.
The effective resolution (ER) due to this effect is defined in terms of the full-scale range of the DAC to the RMS (root-mean-square) residual ripple (both in units of power delivered to the heater, evaluated after the thermal lowpassing effect):
\begin{equation}
    \text{ER} = \log_2 \left( \frac{ P_\text{full-scale} }{ \sigma_\text{ripple} } \right)
\end{equation}

It is possible to analytically estimate the residual ripple and ER for a delta-sigma modulator with large sinusoidal inputs using linearized models that assume uniformly distributed quantization noise with a white noise spectrum.
These assumptions break down for DC inputs (like in a settled control loop) because the delta-sigma modulator produces a periodic output pulse pattern, resulting in a quantization noise spectrum consisting of periodic ``idle'' tones rather than white noise \cite{temes1997delta}.
Instead, we used Simulink to create a time-domain, cycle-accurate model of the delta-sigma modulator followed by a first-order lowpass filter representing the microring's thermal time constant to numerically calculate the worst-case code-dependent residual ripple for varying choices of $f_\text{clk}$ and $\tau_\text{th}$.
The $\sigma_\text{ripple}$ versus DAC code relationship for a given choice of $f_\text{clk}$ and $\tau_\text{th}$ is plotted in Fig. \ref{fig_dac_sim}(b), showing how the worst-case $\sigma_\text{ripple}$ is calculated for a given design point.
The ER trends are plotted in Fig. \ref{fig_dac_sim}(c), also showing $\tau_\text{th}$ reported in previous work in the same platform for high-speed microring modulators with similar geometry to our SFWM microring ($\sim 500\,\text{ns}$ in \cite{decea2019powerhandling}, $2.7\,\text{\textmu s}$ in \cite{mehta2020laserforwarded}, and $14.7\,\text{\textmu s}$ in \cite{sun2016bitstatistics} --- note that $\tau_\text{th} = 1 / ( 2 \pi f_\text{th} ) = 1 / \gamma_\text{th}$).
We found that a high-speed clock frequency of $500\,\text{MHz}$ can support an ER of at least $\approx 10$ bits for the most conservative $\tau_\text{th}$ literature estimate and selected this as our nominal design point (though timing within the block is met up to $f_\text{clk} \sim 1\,\text{GHz}$).
Based on typical $\tau_\text{th}$ reported for undercut microring structures in silicon photonics we expect the true value of $\tau_\text{th}$ to be on the order of $10\,\text{\textmu s}$, leaving a wide margin of safety for the ER and room for future improvement in the number of DAC bits.

\begin{figure}[t!]
    \centering
    \includegraphics[width=\textwidth]{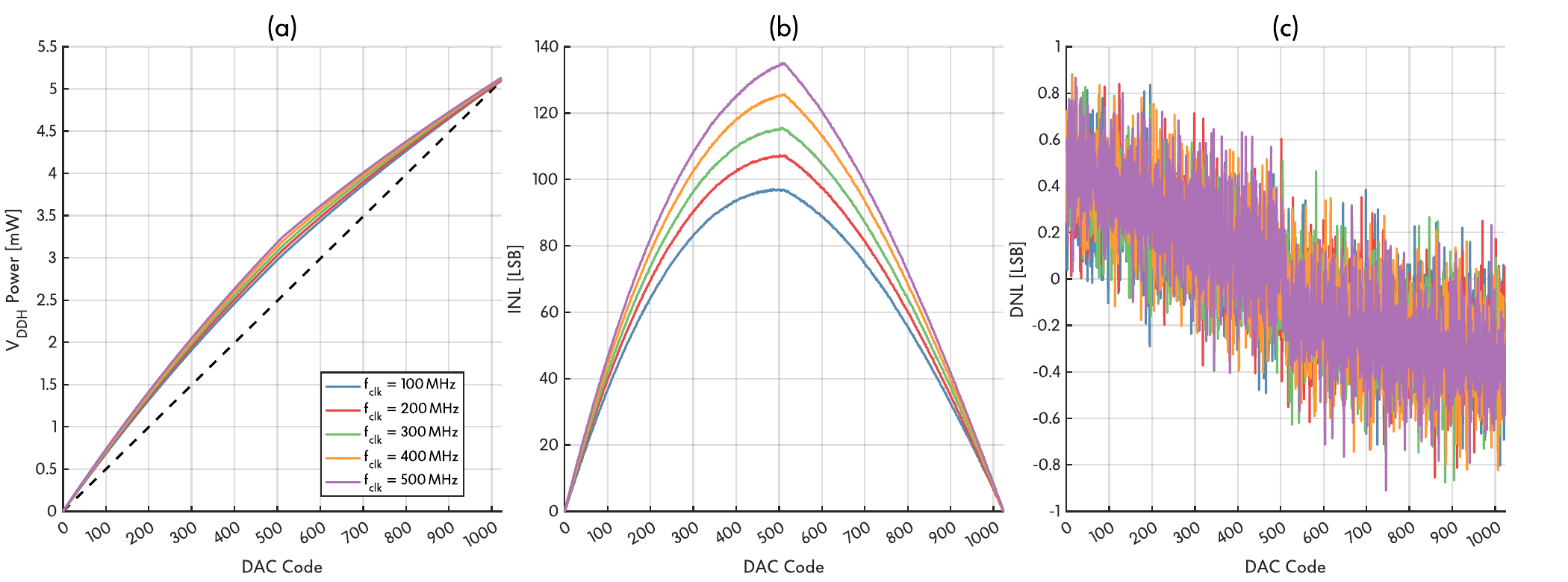}
    \vspace{-1em}
    \caption{(a) Measured DAC code to heater power transfer curve for varying delta-sigma modulator clock frequencies. (b) Integral nonlinearity (INL), or deviation from ideal linear relationship. (c) Differential nonlinearity (DNL), or deviation of each step from ideal LSB size. $\text{DNL} > -1$ indicates monotonicity.}
    \label{fig_dac_char}
\end{figure}

Fig. \ref{fig_dac_char} shows a DC characterization of the heater DAC performed by monitoring the $V_\text{DDH}$ power supply current with varying DAC codes for a single SFWM microring.
The ideal relationship between DAC code and heater power (dashed black line) is linear because the switch-mode driver controls the fraction of time that the heater is fully turned on, rather than the heater current or voltage as would a linear-mode driver (which exhibits quadratic $I^2 R$ or $V^2/R$ nonlinearity between DAC code and heater power).
Slight deviations from a linear response occur due to the uneven rise and fall times in the switching driver circuit --- the rise time constant is $\tau_\text{rise} = R_\text{heater} \, C_\text{parasitic}$ while the fall time constant is $\tau_\text{fall} = (R_\text{switch} || R_\text{heater}) \, C_\text{parasitic}$.
We typically set $R_\text{switch} \ll R_\text{heater}$ to minimize power dissipated in the switch and create a high-efficiency driver, especially in a large photonic system layout like ours with relatively long wires (hundreds of \textmu m) dominating the parasitic capacitance.
The much slower rise time as compared to fall time leads to a net increase in the amount of power dissipated in the heater relative to the expectation for a single pulse from the delta-sigma modulator.
This pushes the transfer characteristic up, especially near midcode where the switching activity factor is greatest, and the effect is stronger at higher clock frequencies because there is less time for the $RC$ response to settle.
The resulting integral nonlinearity (INL) curve is plotted in Fig. \ref{fig_dac_char}(b), while the differential nonlinearity (DNL) curve is plotted in Fig. \ref{fig_dac_char}(c).
For our feedback control application, the INL creates small changes in the loop gain depending on the DAC setting.
This effect barely affects stability margins and is anyway swamped out by the nonlinear response of the microring, so we don't need to minimize INL, preferring to keep the switch-mode driver circuit as simple and compact as possible.
The arithmetic nature of the delta-sigma modulator ensures monotonicity (i.e. $\text{DNL} > -1\,\text{bit}$ across all DAC codes), which is important because the DAC stepping in the wrong direction would reverse the sign of the feedback at that point and cause instability, potentially resulting in a failure to lock or loss of lock.
The measured data confirm this requirement is met.

\subsection{Improving Heater DAC Tuning Range}

The tuning range of our photon-pair source in 45RFSOI is too small to guarantee that fabrication induced variations in resonance wavelength can be compensated across all of the measured dies reported in Table \ref{table:source_ring_variability}.
Here, we discuss issues that contribute to this limited tuning range and show how it can be improved.

\begin{figure}[b!]
    \centering
    \includegraphics[width=\textwidth]{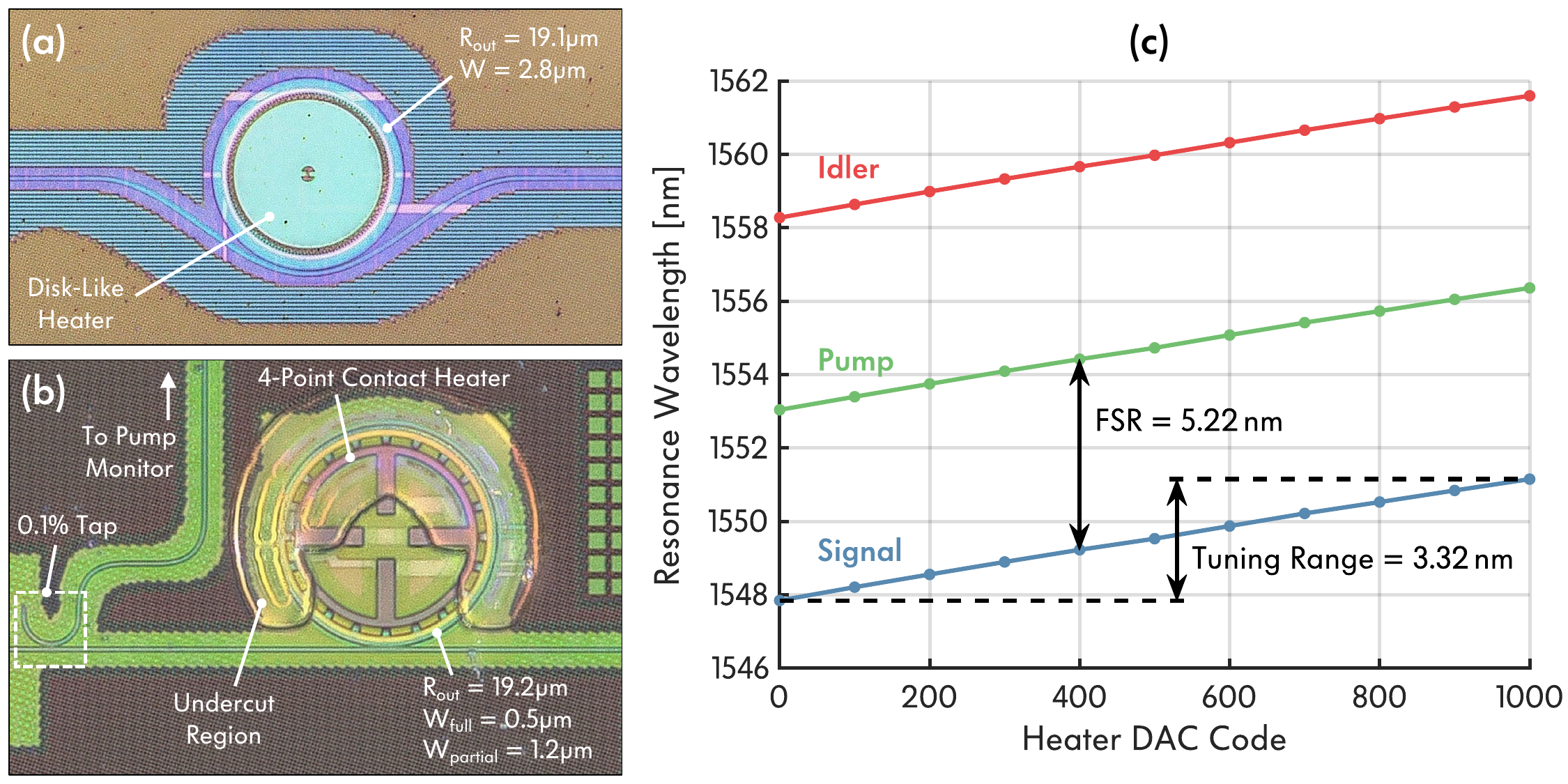}
    \caption{(a) Micrograph of SFWM photon-pair source in 45RFSOI used in our photon-pair generation experiments showing the disk-like heater that results in limited tuning range. (b) Micrograph of SFWM photon-pair source in 45SPCLO (substrate released only for illustration purposes) with improved 4-point contact heater design. The undercut region is visible as an imprint that the TMAH etch leaves on the thick BOX layer. (c) Thermal tuning range of 45SPCLO microring with integrated 1.8\,V heater driver and intact silicon handle substrate. The maximum drive current is 12.5\,mA for a total tuning power of 22.5\,mW to achieve 3.32\,nm of shift. This covers 64\,\% of an FSR, which is sufficient to cover the resonance spread measured in our 45RFSOI design in Table \ref{table:source_ring_variability}.}
    \label{fig_clo_tuning_range}
\end{figure}

The optical phase shift of a thermo-optic phase shifter is determined by the temperature rise it creates in a silicon waveguide, which is proportional to the power dissipated in the heater.
The same heater power can be achieved by driving a high-resistance strip of doped silicon with a higher voltage or a low-resistance strip of doped silicon with a lower voltage (keeping $P$ fixed in $P = V^2/R$).
In principle, this means the power that can be delivered to a heater is not limited by the voltage swing of the driver so long as the heater resistance can be designed appropriately -- and while the voltage rating of the thick-oxide, 160\,nm gate length transistors used to drive the heaters is limited to $1.8\,\text{V} \pm 10\%$, the current rating can be set as high as needed by increasing the total width of the output transistor.

Since we already use heavily doped and silicided silicon to minimize the sheet resistance of the heaters, we must rely primarily on heater geometry and contact placement to tune the resistance (\textit{n}-type doping is used for heaters to minimize sheet resistance because electrons have greater mobility than holes).
In this particular chip tapeout, we had a bug in our heater layout code that limited the tuning range of the microrings -- by targeting a low heater resistance to increase $V^2/R$ with limited $V$, we ended up with the design shown in Fig. \ref{fig_clo_tuning_range}(a) where the heater is too wide and most of the current flows in a loop close to the contacts in the middle of the microring instead of near the silicon waveguide, reducing the waveguide heating efficiency.
This bug in our layout code also forced us to operate with reduced $V_\text{VDDH}$ voltage because the heater resistance was lower than intended and the higher current at full voltage can overheat and damage the output transistor.
A better approach is to limit the width of the heater and add more contacts along its circumference.
This allows a lower resistance to be achieved with a narrow heater by connecting different segments in parallel while keeping the generated heat as close to the waveguide as possible.
It also makes the resistance easier to accurately predict.

Fig. \ref{fig_clo_tuning_range}(b) shows a micrograph of a photon-pair source in 45SPCLO implementing this approach using a 4-point contact heater.
It achieves a greatly improved tuning range of 3.32\,nm, or $64\,\%$ of its FSR, as shown in Fig. \ref{fig_clo_tuning_range}(c).
In future design revisions this tuning range can be further improved by increasing the amount of local substrate undercut \cite{giewont2019globalwgprocess, pal2022low} so that it suspends the entire microring waveguide.
In this chip we were more conservative and chose not to extend the undercut region under the ring-bus coupler to reduce the chances of collapse as the 45SPCLO process was still under development at the time of the tapeout, but finalized design rules that prevent this are now available from the foundry.
These design techniques are also readily applicable to other devices such as Mach-Zehnder interferometers, where the geometric constraints are typically less stringent than in microrings.

The larger heater currents required when the driver voltage swing is limited can still create other issues on the chip such as reduced lifespan via electromigration and electrical crosstalk via $IR$ droops in the power distribution network of the chip.
For near-term demonstrations of quantum systems-on-chip we do not anticipate the total tuning current to become a limiting factor to system size or performance, but it could become issue in future as the system size grows.
A stacked output driver circuit can be used to increase the maximum $V_\text{VDDH}$ voltage \cite{fatemi2019nonuniform}, allowing the heaters to be resized for lower current.
In SOI CMOS stacking is easily accomplished because the bodies of all of the transistors are already isolated by oxide.
This avoids the body-substrate breakdown issues that can occur in bulk CMOS.
The main issue in stacked driver implementation, however, is ensuring that the design is robust to PVT variations and mismatch in the devices, which has not been thoroughly investigated in previous literature.
Verifying that the fabricated stacked drivers operate within the safe operating area of the transistors is challenging because any transient overvoltages within the stack during switching events are not directly observable.
Therefore, accelerated lifetime testing must be performed to ensure the design is robust.

Finally, we must note that improving the heater tuning efficiency results in a correspondingly larger LSB step size.
This means that the number of bits in the delta-sigma modulator and effective resolution must be increased to maintain the same step size for precise tuning.
Supporting a full FSR of tuning range in a single heater driver with the higher Q-factors of microring resonators in 45SPCLO is challenging.
As an alternative, the heater can be partitioned into a low-resistance coarse tuning segment and high-resistance fine tuning segment, separating the process variation calibration and feedback control functions into two different heater driver circuits.










\section{Discussion of Circuit Area Scaling}

\begin{figure}[t!]
    \centering
    \includegraphics[width=\textwidth]{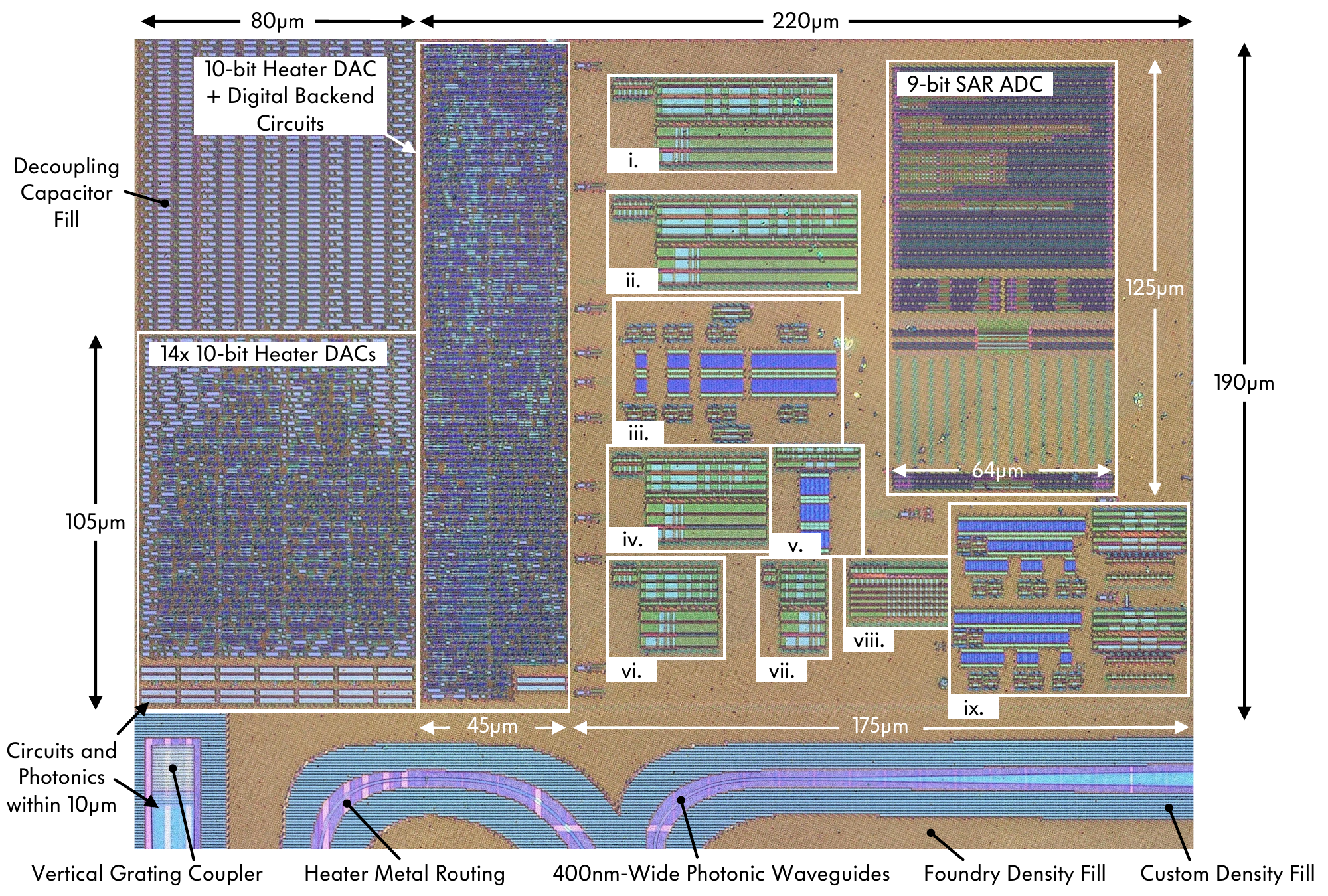}
    \caption{Micrograph of the $220\,\text{\textmu m} \times 190\,\text{\textmu m}$ photocurrent sensor and control circuit block with a detailed area breakdown. An $80\,\text{\textmu m} \times 105\,\text{\textmu m}$ array of 14 heater DACs used for tuning of cascaded-microring filters is also shown on the left side. The numbered sub-blocks of the analog frontend are: i. coarse sense arm offset IDAC (not used), ii. fine sense arm offset IDAC, iii. programmable-gain transimpedance amplifiers, iv. reference arm IDAC, v. ADC driver amplifier, vi. offset IDAC (not used), vii. push IDAC (not used), viii. bias current distribution, ix. comparators and support circuitry for debug (not used).}
    \label{fig_circuit_area}
\end{figure}


\begin{figure}
    \centering
    \includegraphics[width=\textwidth]{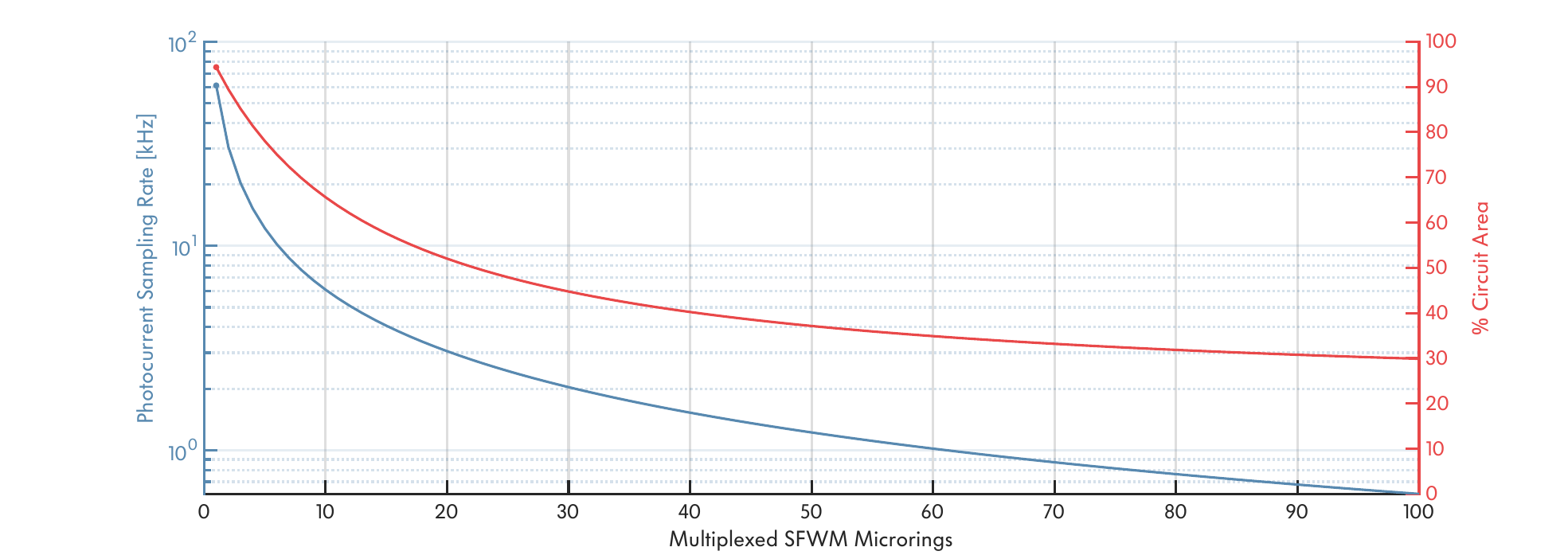}
    \caption{Photocurrent sampling rate and circuit area overhead scaling with the number of multiplexed SFWM microrings per circuit block. The incremental cost of circuit area, a transmission gate and heater DAC, as a fraction of total area per added microring is 21\%.}
    \label{fig_circuit_area_overhead}
\end{figure}

Figure \ref{fig_circuit_area} shows a detailed micrograph of the photocurrent sensor and feedback controller circuit block associated with each SFWM microring on the measured chip, breaking down the area allocated to each subcircuit.
As mentioned in the main text, not all parts of this block need to be duplicated in order to scale-up the number of SFWM microrings that the circuits can control, which reduces the circuit area overhead as the number of SFWM microrings increases.
With robust enough feedback control for stabilizing many SFWM microrings on the same chip achieved at only 10\,Hz update rate in our experiments, the ADC and frontend circuits operating at $\sim$61\,kHz post-averaging (and occupying $175\,\text{\textmu m} \times 190\,\text{\textmu m}$ area) can be time-multiplexed for controlling many microrings on the same chip.
In this case a compact analog multiplexer using the same transmission gate structure as the programmable-gain TIA (each of which occupies $\sim 6\,\text{\textmu m} \times 9\,\text{\textmu m}$) could be added to cycle through the microring photocurrents one at a time, creating only a small area overhead on the analog side for adding more microrings.


The majority of the circuit area that needs to be duplicated per microring comes from the heater DAC and its associated scan chain write register.
The $45\,\text{\textmu m} \times 190\,\text{\textmu m}$ digital circuit area on the left side of the photocurrent sensor contains more than just a heater DAC -- it also includes $512 \times$ digital averaging of the ADC code and an on-chip digital controller circuit block that was bypassed.
Instead, an adjacent circuit block occupying $80\,\text{\textmu m} \times 105\,\text{\textmu m}$ with 14 heater drivers used for cascaded-microring filter tuning provides a better reference for area scaling of the DACs.
Fig. \ref{fig_circuit_area_overhead} shows how the photocurrent sampling rate and percentage of area occupied by on-chip circuits per controlled SFWM microring scales with the total number of multiplexed microrings.
We optimistically assume each microring occupies a footprint of only $50\,\text{\textmu m} \times 50\,\text{\textmu m}$ for the purposes of this calculation, adding 5\,\textmu m margin around the $\sim$20\,\textmu m microring radius to account for the ring-bus coupler, waveguide routing, and density fill.
Even so, the incremental circuit area is only 21\% for each additional multiplexed microring, and the circuits would occupy only 30\% of the area in a system with 100 SFWM microrings sampled at $\sim 610$\,Hz.
Improved packing density and deletion of unused blocks within the analog frontend could reduce this overhead further --- optimizing the circuit area to improve the density of quantum-photonic systems is an interesting direction for future work.
Overall, the trends from this analysis show that monolithic integration compares favorably to heterogeneous integration in spite of electronics and photonics occupying the same plane because the photonics area will be the more-limiting of the two in future large-scale systems.

\section{CMOS Photonics SNSPD and Cryogenic Compatibility}

Integrating SNSPDs with silicon photonics using CMOS-compatible materials and processing steps can make it possible to build scalable quantum-photonic systems with quantum state generation, manipulation, and readout contained on the same chip.
The thin BOX layer of the 45RFSOI CMOS photonics platform allows SNSPDs to be added using transfer printing methods \cite{najafi2015chip, tao2024single}, with light evanescently coupling from the waveguide to nanowires attached to the back side of the die after the XeF$_2$ etch step.
The thick BOX in the next-generation 45SPCLO electronic-photonic process can also be locally etched to directly expose the silicon waveguide layer and integrate devices and materials directly onto the photonics post-fabrication \cite{onural2023toward}.
In the long run, however, a better integration approach is to directly pattern nanowires made of materials such as niobium nitride (NbN) \cite{pernice2012snspdsiph}, niobium-titanium nitride (NbTiN) \cite{akhlaghi2015waveguide}, tungsten silicide (WSi) \cite{shainline2017room}, or molybdenum silicide (MoSi) \cite{li2016nano} onto or near the silicon waveguide layer as a custom process module at the CMOS foundry.
PsiQuantum has recently demonstrated the implementation of such a capability in a silicon photonics process from GlobalFoundries \cite{alexander2024manufacturable} including a metal barrier for shielding from stray light, but so far without integrated electronics, and without yet providing details on the specific materials or processing steps used.

Integrating SNSPDs means that the whole chip will need to operate at temperatures on the order of a few Kelvin, which are readily accessible via closed-cycle cryocoolers.
High-performance NbTiN SNSPDs have also recently been engineered to operate as high as 7\,K \cite{gourgues2019superconducting}, promising to ease thermal budget requirements for such systems.
The key technical consideration involved in operating the SFWM sources described here at cryogenic temperatures are the compatibility of the electronics, the effectiveness of thermal tuning of silicon-photonic devices, and the different behavior of silicon microrings at cryogenic temperatures.
Fortunately, competing effects on mobility and threshold voltage produce a net increase in the performance of MOSFETs in 45RFSOI when operated at cryogenic temperatures down to 2.5\,K.
This result was obtained with MOSFET and ring oscillator test structures on the same chip as our quantum-photonic systems \cite{yin2024cryogenic}.
It is consistent with the general literature trend of deep submicron CMOS transistors operating with improved performance for cryo-CMOS applications \cite{charbon2021cryo}, since field-induced ionization from the gate voltage overcomes carrier freezeout effects.
Integration of SNSPDs with high-performance cryo-CMOS circuits can also enable active gating \cite{hummel2022nanosecond}, quenching \cite{ravindran2020active}, and readout signal amplification \cite{peng20240} with ultra-low parasitics, improving their single-photon detection performance metrics.

Given that our system depends on thermal tuning to stabilize microring photon-pair sources, the next important consideration is whether thermal tuning is practical at cryogenic temperatures.
Silicon has a low thermo-optic (TO) coefficient of approximately $10^{-8} \text{\,K}^{-1}$ at deep cryogenic temperatures on the order of a few K, compared to $1.8 \times 10^{-4} \text{\,K}^{-1}$ at 300\,K \cite{komma2012cryo}.
Using mode simulations and the temperature-dependent TO coefficient, we estimate that a 12\,K temperature rise is needed to generate the 0.62\,nm tuning range observed in our microring source near room temperature, while an 89\,K shift is needed to achieve the same range starting at 5\,K (the lowest temperature at which the TO coefficient is measured in \cite{komma2012cryo}).
Degenerately doped and silicided heaters will not suffer from carrier freezeout, but the temperature dependence of their resistance needs to be characterized to target their design point properly.
The local temperature increases at the photon-pair sources must also be carefully isolated from the SNSPDs to ensure their performance is maintained.
To make thermal tuning compatible with SNSPDs at cryogenic temperatures in 45SPCLO, isolation in the form of undercut trenches (as shown earlier in Fig. \ref{fig_clo_tuning_range}(b)) can be used to limit thermal crosstalk, while heatsinking the silicon substrate to the cryogenic cold head to anchor the temperature of the entire die.
This technique was used to enable thermal tuning of Mach-Zehnder interferometers integrated with SNSPDs in the PsiQuantum silicon quantum photonics platform \cite{alexander2024manufacturable}.
In 45RFSOI the substrate is already fully removed, so the die must be anchored to the cold head through the Cu pillar side with vias through the package.
Heat transport simulations and measurements with test structures are required to establish the safe minimum distance between the thermal tuners and SNSPDs.

When considering the challenges of implementing thermally-tuned microring SFWM sources at cryogenic temperatures, it is also important to note the benefits.
The heat capacity \cite{pernice2011carriers} and thermal conductivity \cite{thompson1961thermal, glassbrenner1964thermal} of silicon both decrease at deep cryogenic temperatures, suggesting that less heater power will be required for a given temperature shift.
Another benefit is that the two-photon absorption (TPA) coefficient for 1550\,nm light is halved at cryogenic temperatures while the effect on the $\chi^{(3)}$ nonlinearity responsible for SWFM is weaker \cite{sinclair2019tpa}.
Below 10\,K, the free-carrier lifetime in silicon is also reduced by an order of magnitude in comparison to room temperature \cite{pernice2011carriers}.
Overall, this leads to less free-carrier absorption in silicon microrings at cryogenic temperatures and enables operating at higher pump powers where free carrier effects otherwise lead to saturation of the PGR due to lowered intrinsic Q-factor \cite{xiang2020effects}.
Additionally, the consequent reduction in self-heating pushes up the pump power level where thermal bistability occurs, as was recently demonstrated experimentally \cite{lin2023cryogenic}, making them easier to control and stabilize.
The most important benefit of cryogenic operation, however, is that noise from spontaneous Raman scattering in the amorphous cladding materials is dramatically reduced.
This improves the CAR by more than a factor of 2$\times$ \cite{feng2023entanglement, witt2024packaged} and makes it possible to take advantage of silicon microrings' increased power handling capability.

\section{CMOS Photonics Optical Packaging}

We observe a small amount of drift in the fiber alignment and coupling efficiency in our experiments, forcing us to introduce some additional back-off from the maximum photocurrent in longer-duration measurements.
This occurs because we use manual positioning stages that can slowly creep over time (Thorlabs Nanomax 300 with differential micrometers) to hold the lensed fibers over the chip that couple light in and out of the electro-optical systems.
Similar positioning stages with motors or piezo actuators and closed-loop feedback from built-in sensors could be used in future quantum-photonic system demonstrations to avoid this issue at the prototype stage. 
An alternative option that we have incorporated in next-generation chips is a non-resonant monitor photodiode tapping off a small amount of input light ($<1\,\%$) to compensate for the drift by rescaling the downstream photocurrent readings.

\begin{figure}[t!]
    \includegraphics[width=\textwidth]{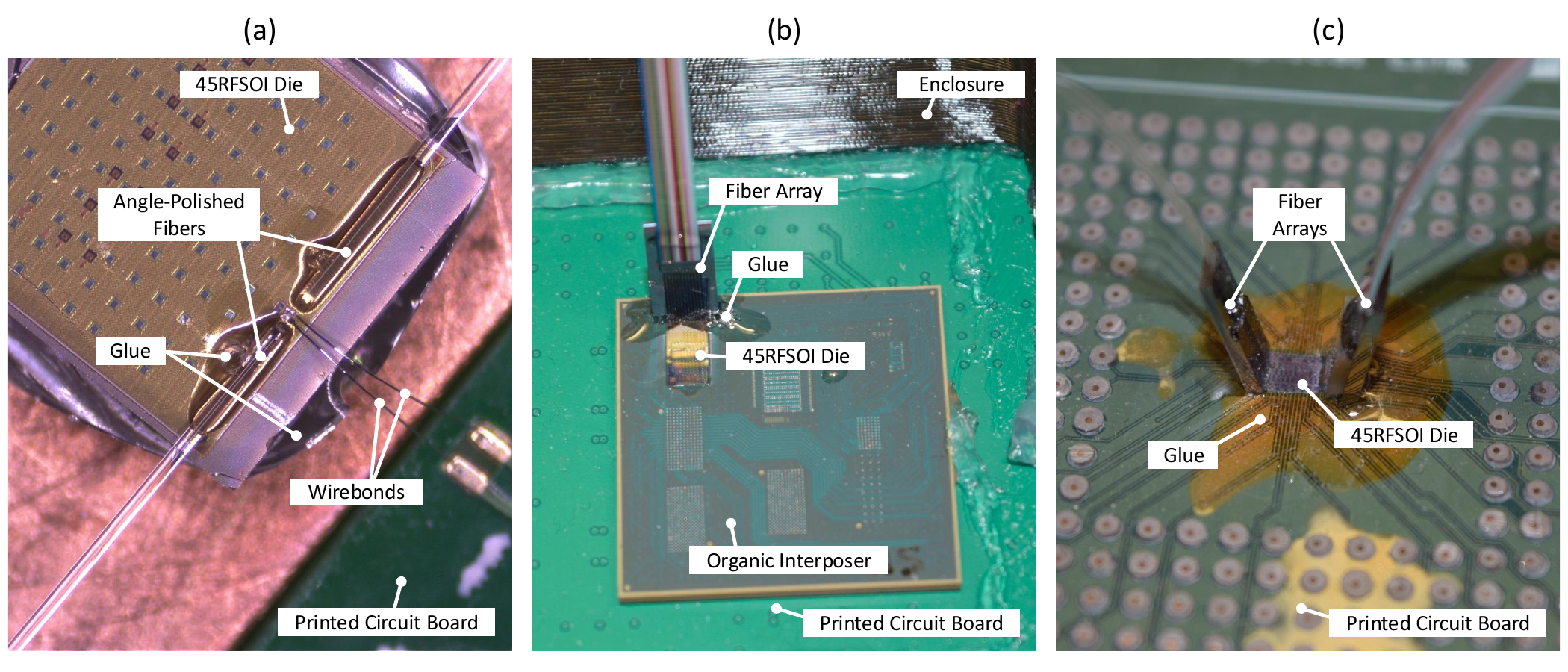}
    \caption{(a) Cryogenic packaging of substrate-transferred 45RFSOI die using angle-polished fiber approach, reproduced with permission from \cite{decea2020photonic}. (b) Fiber array packaging of flip-chip 45RFSOI die for ultrasound sensing \cite{zarkos2022fully}. (c) Dual fiber array packaging of flip-chip 45RFSOI die for analog RF-over-fiber transceiver \cite{buchbinder2023analog}.}
    \label{fig_fiber_packaging}
\end{figure}

In fully-packaged systems, optical fibers are glued to the chip to attach them permanently and avoid the issue of alignment drift during an experiment.
A wide variety of techniques exist to do this \cite{kopp2010silicon, boeuf2015silicon, carroll2016photonic, ranno2022integrated}, which can also function down to cryogenic temperatures needed for SNSPD compatibility when appropriate adhesive materials are used.
Vertical grating couplers (VGCs), like those used in the systems reported here, require active alignment (typically from a loopback measurement) to position and glue an optical fiber or fiber array over the exact location on the chip where the coupling efficiency is maximized.
In the simplest case, individual angle-polished fibers can be glued down flat against the chip surface such that the reflection at the end couples down into a grating coupler \cite{snyder2013packaging}.
This technique does not require any exact spacing between grating couplers like in a fiber array or any additional support structures on the chip to hold the fiber at an angle (e.g. \cite{shainline2017room}), making it possible to implement in the future without any redesign of our system layouts or additional post-processing.
In \cite{mckenna2019cryogenic}, angle polished fibers glued over a bidirectional grating coupler achieve $\sim 5\,$dB coupling loss at room temperature (the theoretical minimum for a bidirectional coupler is 3\,dB) and $\sim 7$\,dB loss after repeated cycling to millikelvin temperatures.
Fig. \ref{fig_fiber_packaging}(a) shows how the same approach was also recently demonstrated in the 45RFSOI CMOS photonics platform used here to optically package a microring modulator used at 3.6\,K for readout of co-packaged SNSPDs \cite{decea2020photonic}, but with high losses (10\,dB at room temperature and 15\,dB at 3.6\,K) due to mismatch between the 5\,\textmu m mode-field diameter (MFD) grating couplers, similar to those used here, and the commonly available 9.2\,\textmu m MFD angle polished SMF28 fiber.
The authors estimate that losses could be reduced down to the few-dB level by redesigning the grating couplers to match the fiber mode and optimizing their response at cryogenic temperatures instead of room temperature.
Design techniques for improving the misalignment sensitivity of VGCs \cite{romero2015alignment} could also be incorporated to reduce the additional insertion loss after cooling to cryogenic temperatures.

A more conventional way to package silicon photonics products with VGCs is to glue a fiber array on a glass block to the surface of the chip, as shown in Fig. \ref{fig_fiber_packaging}(b).
The grating couplers must be placed in a line matching the pitch of the fiber array, with standard pitches being 127\,\textmu m and 250\,\textmu m.
Unlike with individual fibers, back reflections at the chip interface can couple a small amount of light between all of the fibers through the glass block.
In \S 3.6.2 of \cite{oser2019integrated}, the best-case crosstalk within a fiber array was measured to be $-90$\,dB between ports on the opposite ends of a 6-channel array, meaning that a single fiber block is unsuitable for simultaneously delivering the pump and extracting the post-filtered signal and idler photon streams --- at least two separate fiber arrays are required, as shown in Fig. \ref{fig_fiber_packaging}(c), and further work is needed to establish spacing rules to ensure sufficient isolation.
A sizable area ($\sim$mm in each direction) must be reserved for each fiber block with additional clearance for glue, making it impractical for us to implement this packaging scheme within our section of the $2\,\text{mm} \times 9\,\text{mm}$ multi-project chiplet.
A fiber array packaging scheme for 45RFSOI datacom transceiver chiplets integrated on a multi-chip module has previously been demonstrated \cite{sun2020teraphy, wade2021teraphy} and could be developed further with cryogenic-compatible adhesives.
Such an approach for packaging a silicon microring photon-pair source for cryogenic operation has already been demonstrated with minimum insertion losses of around 4.1\,dB achieved after cooling to 5.3\,K, but with the transmission peak occurring at 1620\,nm rather than the intended 1560\,nm wavelength used for photon-pair generation \cite{witt2024packaged}.
In principle the fiber array approach should achieve the same level of performance as individual angle-polished fibers, but with better scalability for increasing the fiber count.

In GlobalFoundries' next-generation 45SPCLO CMOS photonics process \cite{rakowski2020cloprocess}, additional processing steps enable low-loss passive alignment of fiber arrays to V-grooves etched into the edge of the chip.
A metamaterial waveguide section transitions between the $\sim 400\,\text{nm}$-wide single-mode waveguides and 9.2\,\textmu m MFD fiber mode, and $<1\,\text{dB}$ best-case overall insertion loss has been demonstrated during development of this process \cite{barwicz2018integrated} with 2\,dB being typical across wavelength, alignment, and fabrication process variations.
A cryo-compatible edge coupling scheme based on glued fibers without V-grooves also achieved similar loss values in recent work on delivering light for manipulation of trapped-ion qubits on a photonic integrated circuit operating at 7\,K \cite{mehta2020integrated}.
The improved mechanical support afforded by the V-grooves etched in silicon makes this approach appealing for operating over wide temperature ranges, but the drawback is large area consumption for the V-grooves and transitions as compared to grating couplers, where the glass fiber block does not prevent circuits or photonics from being placed underneath.
Grating couplers with similarly low $<1\,\text{dB}$ minimium loss (but without any study yet of variability) have also recently been demonstrated in this platform \cite{zhang2022sub}, making both approaches promising candidates for future testing with cryogenic-compatible adhesives.
Some other approaches not yet standard at silicon photonics foundries, such as photonic wirebonds \cite{lin2023cryogenic}, also show promise for cryogenic packaging.

\newpage
\bibliography{references_supplementary}